\documentclass[a4paper,10pt]{article}
\pdfoutput=1 
\usepackage{jcappub}
\usepackage[T1]{fontenc}
\usepackage{appendix}
\usepackage{enumitem}
\usepackage{booktabs}
\usepackage{multirow}
\usepackage{siunitx}
\usepackage{orcidlink}
\usepackage{arydshln}
\usepackage{caption}
\usepackage{subfig}
\usepackage{graphicx}

\title{\boldmath Improving data-driven model-independent reconstructions and updated constraints on dark energy models from Horndeski cosmology}

\author[a]{Mauricio Reyes}
\author[a]{and Celia Escamilla-Rivera\orcidlink{0000-0002-8929-250X}}

\affiliation[a]{Instituto de Ciencias Nucleares, Universidad Nacional Aut\'onoma de M\'exico, Circuito Exterior C.U., A.P. 70-543, M\'exico D.F. 04510, M\'exico.}
\emailAdd{mauricio.cruz@correo.nucleares.unam.mx}
\emailAdd{celia.escamilla@nucleares.unam.mx}

\abstract
{In light of the statistical performance of cosmological observations, in this work we present an improvement on the Gaussian reconstruction of the Hubble parameter data $H(z)$ from Cosmic Chronometers, Supernovae Type Ia and Clustering Galaxies in a model-independent way in order to use them to study new constraints in the Horndeski theory of gravity. First, we have found that the prior used to calibrate the Pantheon supernovae data significantly affects the reconstructions, leading to a 13$\sigma $ tension on the $H_0$ value. Second, according to the $\chi^{2}$-statistics, the reconstruction carried out by the Pantheon data calibrated using the $H_{0} $ value measured by The Carnegie-Chicago Hubble Program is the reconstruction which fits best the observations of Cosmic Chronometers and Clustering of Galaxies datasets. Finally, we use our reconstructions of $H(z)$ to impose model-independent constraints in dark energy scenarios as Quintessence and K-essence from general cosmological viable Horndeski models,  landscape in where we found that a Horndeski model of the K-essence type  can reproduce the reconstructions of the late expansion of the universe within 2$\sigma$.
}

\begin{document}
\maketitle
\flushbottom

%%%%%%%%%%%%%%%%%%%%%%%%%%%%%%%%%%%%%%%%%%%%
%%%%%%%%%%%%%%%%%%%%%%%%%%%%%%%%%%%%%%%%%%%%

\section{Introduction}
\label{sec:intro}
Currently,  local measurements of the Hubble-Lemaître parameter from supernovae type Ia \cite{Riess:2016jrr,Riess:2018byc} and lensing time delays \cite{Birrer:2018vtm} are in disagreement with the value inferred from a $\Lambda$CDM~fit to the Cosmic Microwave Background (CMB) \cite{Aghanim:2018eyx}. This discrepancy has not been easily explained by any obvious systematic effect \cite{Aghanim:2016sns,Aylor:2018drw} in either measurement. According to this, an increasing discussion on the cosmology community \cite{DiValentino:2021izs,DiValentino:2020zio}  is focusing on the possibility that this \textit{Hubble-Lemaître tension} may be indicating new physics beyond the concordance model \cite{Freedman:2017yms}. 

However, theoretical proposals for the Hubble-Lemaître tension are not easily within reach, since
one of the biggest challenge remains in the calculations of the angular scale of the acoustic peaks in the
CMB power spectrum, which fix the ratio of the sound horizon at decoupling to the distance to the CMB surface of last scatter.  Furthermore, some solutions 
involve early-time modifications on the theory of gravity itself, which change the sound horizon, or even late-time changes to the expansion rate.  
On the side of  late-time proposals, we can mention: a phantom-like dark energy (DE) component \cite{DiValentino:2017zyq}, a vacuum phase transition \cite{Banihashemi:2018has}, or interacting DE \cite{Kumar:2016zpg}. Moreover, these models are tightly constrained \cite{Addison:2017fdm,DiValentino:2017iww} by late-time observables, especially those from baryon acoustic oscillations (BAO) \cite{Alam:2016hwk}. Model-independent parameterizations of the late-time expansion history are similarly constrained \cite{Zhao:2017cud}, while others proposals aim to put theoretical background to parameterizations modified gravity at the same level as other parameterizations  into the pipeline and analysis of observational data and forecasts \cite{Jaime:2018ftn}.  On the early-time resolution, it is possible to reduce the sound horizon with radiation energy density, which can be constrained by BAO and by the CMB power spectrum \cite{DiValentino:2017iww}. As an extension, it is also possible to address the Hubble tension through a direct modification of gravity \cite{DiValentino:2017iww,DiValentino:2020zio} that is active around late and early times along the Hubble flow, e.g. in Galileon models \cite{Renk_2017,Peirone_2018,Peirone:2019aua,Frusciante_2020} and also in extensions as Teleparallel Gravity \cite{Nunes:2018xbm,Escamilla-Rivera:2019ulu,Rave-Franco:2021yvu}. Furthermore, in \cite{Raveri:2019mxg} was shown a method to reconstruct  Horndeski cosmological models using the Effective Field Theory approach \cite{Gubitosi:2012hu} and interpolation methods with a  piece-wise fifth order spline.

The discussed Hubble tension is one of the several issues\footnote{Despite great efforts, dark matter remains undetected and the nature of the Cosmological Constant via dark energy continues to have several issued associated with it.}   why  
it has recently become of interest the  study of modified theories of General Relativity (GR). It has been of particular interest the study of   Hordesnki theory \cite{bellini2020hi_class,ballesteros2020h0,emond2020black}, this is the most general scalar-tensor theory in four dimensions that is built only from the metric tensor $g_{\mu\nu}$  and a scalar field $\phi$ that is able of avoiding Ostrogradsky instabilities \cite{kobayashi2019horndeski}. The theory has four free functions $G_{2},G_{3},G_{4},G_{5}$ that depend on the scalar field $\phi$ and its kinetic term $2X=-\partial_{\mu} \phi\partial^{\mu}\phi$. The correct choice of the $G_{i}$ functions allows us to recover various theories including models such as Quintessence, Brans–Dicke, k-essence and others \cite{kase2019dark}. At the present time observations \cite{kobayashi2019horndeski} have reduced the theory to $G_5 = 0$ and $G_4$ to be   a function that just depends on the scalar field $G_4(\phi)$. Given  the generality of this theory, if the tension $ H_ {0} $ can be solved by resorting to an extension to GR, Horndeski theory will be able to describe that extension, at least effectively with a proper choice of $G_{i}$. In this work it will be of particular interest to find a combination of $G_{i}$'s that could be a candidate to solve the $H_{0}$ tension \cite{Zumalacarregui:2020cjh}.

Furthermore, when consider data from a specific survey we have to deal with a problem of model-dependency according to the nature of the observations, e.g. observations from CMB are biased by a $\Lambda$CDM model. This problem has motivated the study of other data treatment methods that extract important information directly from the observational data, as for example promising alternatives related to non-parametric reconstructions of the cosmic expansion \cite{Montiel:2014fpa, mehrabi2020does,Briffa:2020qli,gomez2018h0}.
These kinds of approaches attempt to reconstruct the cosmological evolution directly from observations without establishing an association with a theoretical physical model. In this line of thought, we can divide these approaches as model-dependent and model-independent methods. Model-dependent methods are useful when the relationship between the variables of the landscape under study is known, and their objective is to constrain the parameters of the chosen cosmological model. Moreover, when there is not any hint about the explicit form of this relationship we need to propose it \textit{at hand}, which can leads to biased results. On the other hand, model-independent methods provide a generic trend of the parameter of interest when the relationship between the parameters is unknown or there is little feedback about it. These kinds of methods have become very recurrent given their characteristics for enhancing scatter and other diagnostic analyses whose can display the underlying structure in the data. Some example in that regard is the Gaussian Process (GP) \cite{seikel2012reconstruction}, which is a Bayesian statistical approach that allows to obtain a function that describes a specific data set without an assuming a particular cosmological model. But of course, the price we pay to have a \textit{model-independent} method is the need of a  prior for the function that we seek to reconstruct and a likelihood for the observational data. Once this information is given we can calculate the posterior distribution for the functions that can describe the observations.

According to the ideas discussed above, we will study the late-time expansion of the universe in a model-independent way through the reconstruction of the late observational data. To perform this analysis, we will make use of observational data that involve an improved GP reconstruction version of $H(z)$. Once we have at hand the reconstructed late expansion of the universe, we proceed to use our new reconstructions to found new cosmological constraints in Hordenski models, in particular we will study a Quintessence  and  a K-essence model and develop a possible solution of $H_0$ tension at late-times. In this line of research, interesting analyses have been done in \cite{Zumalacarregui:2020cjh} where it was found that early modified gravity techniques can reconcile the $H_0$ value by increasing the expansion rate in the epoch of matter-radiation equality. Also extensions up to cubic covariant Galileon model has been study in \cite{Frusciante:2019puu}.
Furthermore, surviving Horndeski effective field theory of dark energy has been treated at all redshift with the Gravitational Waves constraints \cite{Frusciante_2019}.

We outlined this paper as follows: in Sec. \ref{sec:recons} we explain our new GP methodology on the observational datasets to perform model-independent reconstructions of $H(z)$. Also, we include the description of GP \textit{kernels} used to perform such reconstructions. In Sec. \ref{sec:model_constraints} we consider a model-independent methodology to constraint dark energy density at late-times showing that it is possible to report a 1$\sigma$ statistical uncertainty with a numerical error associated with the calculation of this density. 
In Sec. \ref{sec:Horn} we employ these model-independent reconstructions to derive new constraints on specific relations of the Galileon  $G_2$  from Horndeski theory of gravity. We found that under certain initial conditions our system can be reduced to a one-parameter to be fitted with the observations. 
Finally, we present our results and conclusions in Sec. \ref{sec:conclu}.

%%%%%%%%%%%%%%%%%%%%%%%%%%%%%%%%%%%%%%%%%%%%
%%%%%%%%%%%%%%%%%%%%%%%%%%%%%%%%%%%%%%%%%%%%

\section{Improving GP reconstructions for the late-time data}  
\label{sec:recons}

Gaussian Processes (GP) employed in cosmology provide a method to extract information from observations without assuming a particular model for the description of the data, which makes them a powerful tool, since they can help us to understand the nature of the late time expansion of the universe \cite{seikel2012reconstruction,williams2006gaussian}.  From a statistical point of view, GP is a stochastic process formed by a collection of random variables $\boldsymbol{x}$, such that every finite linear combination of these variables has a multivariate Gaussian distribution
\begin{equation}
    \boldsymbol{x}\sim  \mathcal{N}(\mu,\Sigma),
\end{equation}
with $\mu$ being the vector that contains the mean value of the random variables and $\Sigma$ the covariance matrix between the random variables. A Gaussian process for a function $f(z)$ is described by a covariance function (\textit{kernel}) $k(\widetilde {z}, z)$  and by its mean value. The covariance function expresses the fact that the  function  evaluated at one point $z$ is not independent of the  function evaluated at other point $ \widetilde {z} $, but rather these are correlated and the relationship is given by the covariance function. There is a wide range of possible covariance functions\footnote { As long as a reasonable covariance function is chosen for the problem at hand. In our case we need the covariance function to be continuous and  differentiable at least one time. Also, we expect that the correlation between the function evaluated at two points decreases  with the distance between them.}, moreover it has been shown that the choice of a kernel does not significantly influence the derived value of the cosmological parameters to be determined by the reconstruction \cite{Briffa:2020qli,gomez2018h0,mehrabi2020does}, i.e. even when the mean value of these parameters varies when we change the covariance function, these values coincide within 1$\sigma $. This motivates us to consider the following two different kernels for our analysis:
\begin{eqnarray}
 k(z,z')&=&\sigma_{f}^{2} \exp\left (  \frac{-(z-z')^2}{2l^{2}}\right ),  \quad \rightarrow \quad \text{Squared Exponential} \label{eq:SE} \\
 k(z,z')&=&\sigma_{f}^{2} \left (  \frac{l_{f}}{     (z-z')^{2}     +l_{f}^{2}} \right ), \quad \rightarrow \quad \text{Cauchy function} \label{eq:CF}
\end{eqnarray}
where $\sigma_{f}$ determines the width of the reconstructed function  and $l$ determines how  the correlation between two points change, a large $l$  results in a smooth reconstructed function, while a small $l$ gives an oscillating function \cite{seikel2012reconstruction}.

GP have been widely used in the literature for different  purposes, ranging from the  study of  growth rate of structure $f \sigma_{8}(z)$ \cite{zhang2018gaussian}, to  obtain  constraints  on Teleparallel gravity models \cite{Briffa:2020qli}, and even some authors have  found a way to  use the reconstructions to test the $\Lambda$CDM model, in   \cite{seikel2012reconstruction} 
they  use the reconstructions of $H(z)$ to look for deviations from the equation of state of the cosmological constant $\Lambda $,  in \cite{seikel2012using} they found a way to test the concordance model  using a diagnostic function $ O_{m} $ \cite{seikel2012using,escamilla2016nonparametric}. 
To proceed our reconstructions we require the following samples:

\begin{enumerate}
\item \textbf{Pantheon SNeIa sample.} 
Consists of 1048 SNeIa in 40 bins compressed. Currently, this sample is the largest spectroscopically supernovae Ia data set. These astrophysical objects can give determinations of the distance modulus $\mu$,  its theoretical prediction is related to the luminosity distance $d_L$
\begin{equation}\label{eq:lum}
\mu(z)= 5\log{\left[\frac{d_L (z)}{1 \text{Mpc}}\right]} +25,
\end{equation}
where the luminosity distance is given in Mpc. In this expression we need to add the nuisance parameter $M$, as an unknown offset sum of the supernovae absolute magnitude (with other possible systematics), which can be degenerate with the $H_0$ value. We assume
a spatial flatness and $d_L$ can be related to the comoving distance $D$ using
$d_{L} (z) =\frac{c}{H_0} (1+z)D(z),$
where $c$ is the speed of light, and obtain
\begin{equation}
D(z) =\frac{H_0}{c}(1+z)^{-1}10^{\frac{\mu(z)}{5}-5}.
\end{equation}
The normalised Hubble function $H(z)/H_0$ is derived from the inverse of the derivative of $D(z)$ with respect to $z$
$D(z)=\int^{z}_{0} H_0 d\tilde{z}/H(\tilde{z}), $
where $H_0$ is the Hubble value consider as a prior  to normalise $D(z)$. For our sample, we calibrated this data 
by using four different priors for $H_0$:

\begin{itemize}
    \item  Cefeids \cite{Riess:2016jrr} :   $H_ {0}^{\text{R}} = 73.24 \pm 1.74 \text{km/s/Mpc}$. 
    \item  Planck 2018 \cite{Aghanim:2018eyx}:  $H_ {0}^{\text{Planck}} = 67.4  \pm 0.5 \text{km/s/Mpc}$. 
    \item  H0LiCOW \cite{wong2020h0licow}: $H_ {0}^{\text{HW}} = 73.3^{+1.7}_{- 1.8}\text{km/s/Mpc}$. 
    \item  TRGB \cite{freedman2019carnegie}:    $H_{0}^{\text{TRGB}} = 69.8 \pm 1.9\text{km/s/Mpc}$. 
\end{itemize}

\item \textbf{Cosmic Chronometers sample (CC)}.  This data  provide    model independent measurements for H(z). These measurements are obtained by comparing the stellar evolution of galaxies with low star formation and similar metallicity \cite{jimenez2002constraining}.  Currently there are 31 model-independent measurements \cite{magana2018cardassian} of $ H(z)$ obtained through this method.

\item \textbf{Clustering Galaxies sample (CL)}. 
Usually, this sample  analyse  the baryon acoustic oscillations (BAO) with   a two-point correlation function,  the analysis allows them  to get   measurements \cite{wang2017clustering} of the product $H (z)r_{d}$, where $ r_ {d} $ is the   sound horizon  at the  drag epoch. If we assume a fiducial  cosmology they can   estimate  $r_{d}$, and then break the degeneracy between $r_{d}$ and $H(z)$. Currently, there are  20 measurements of $H (z)$ obtained with this method  \cite{magana2018cardassian}. However, we must be careful working with this data, since is biased due to an underlying $\Lambda$CDM cosmology. 
\end{enumerate}

\begin{figure*}
    \centering
    \includegraphics[scale=0.247]{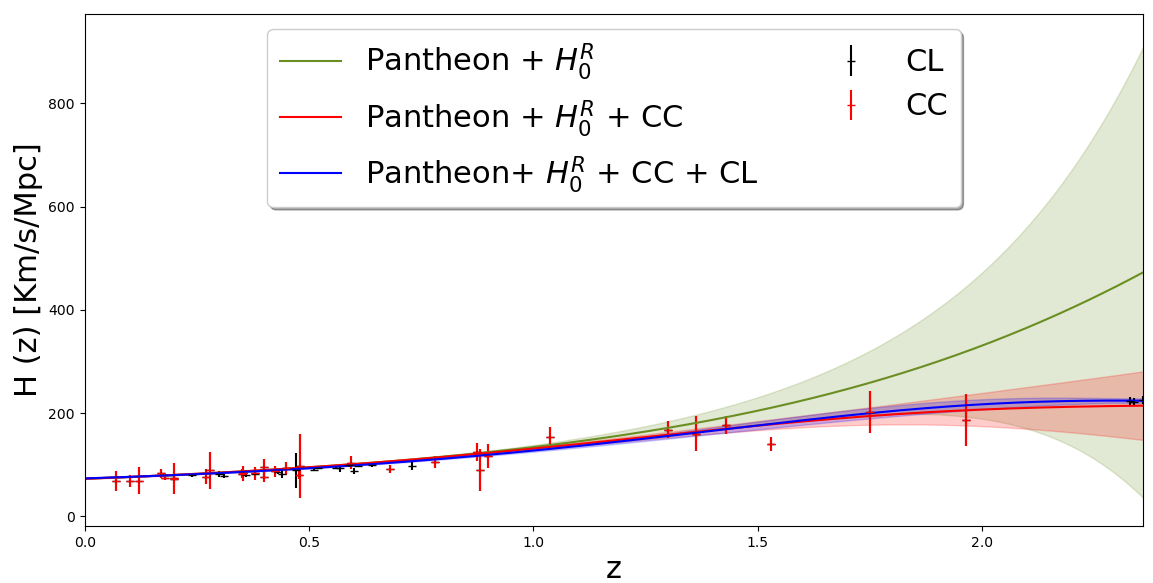} 
\includegraphics[scale=0.247]{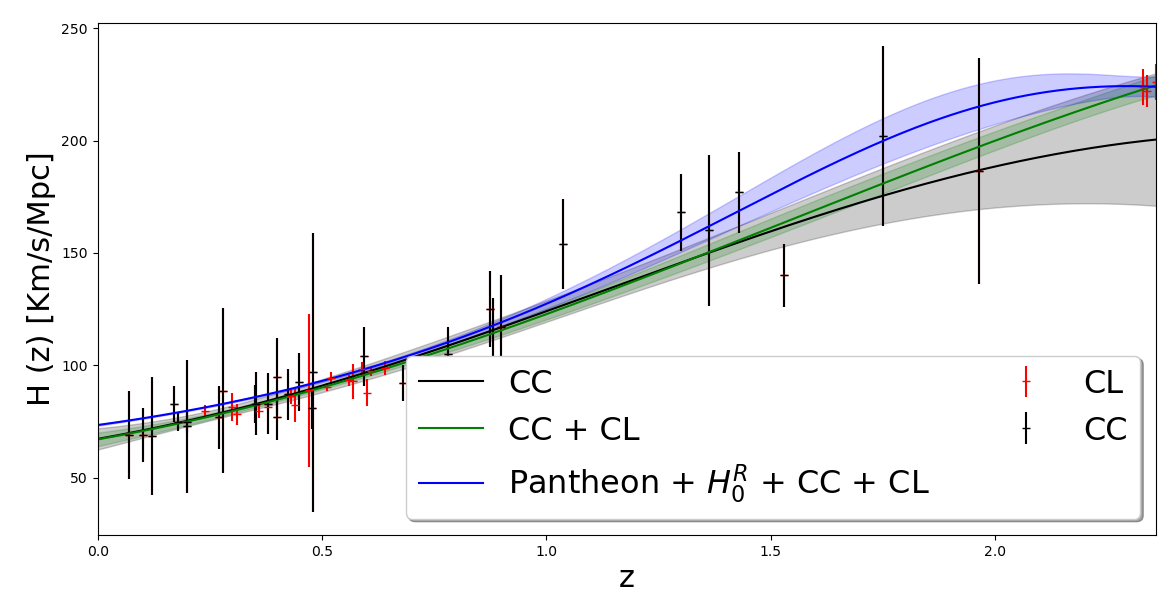} 
\includegraphics[scale=0.21]{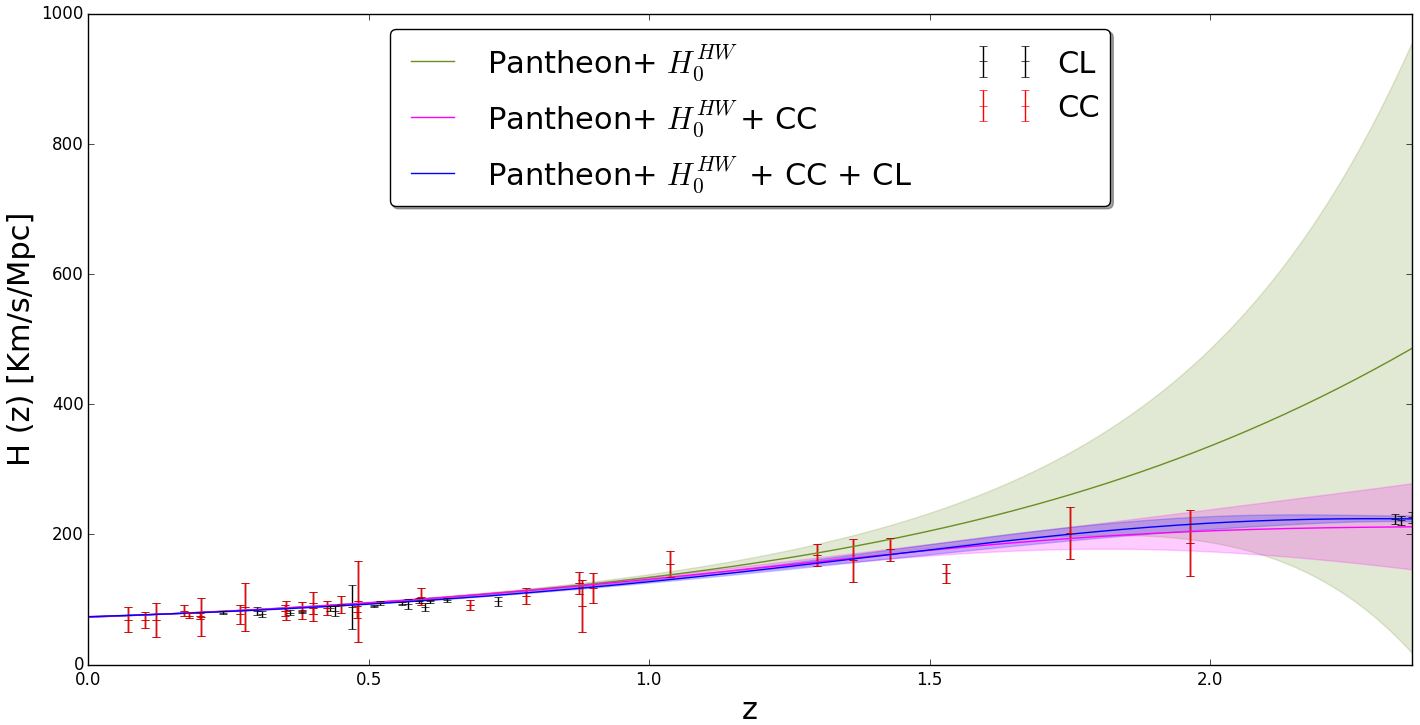} 
\includegraphics[scale=0.21]{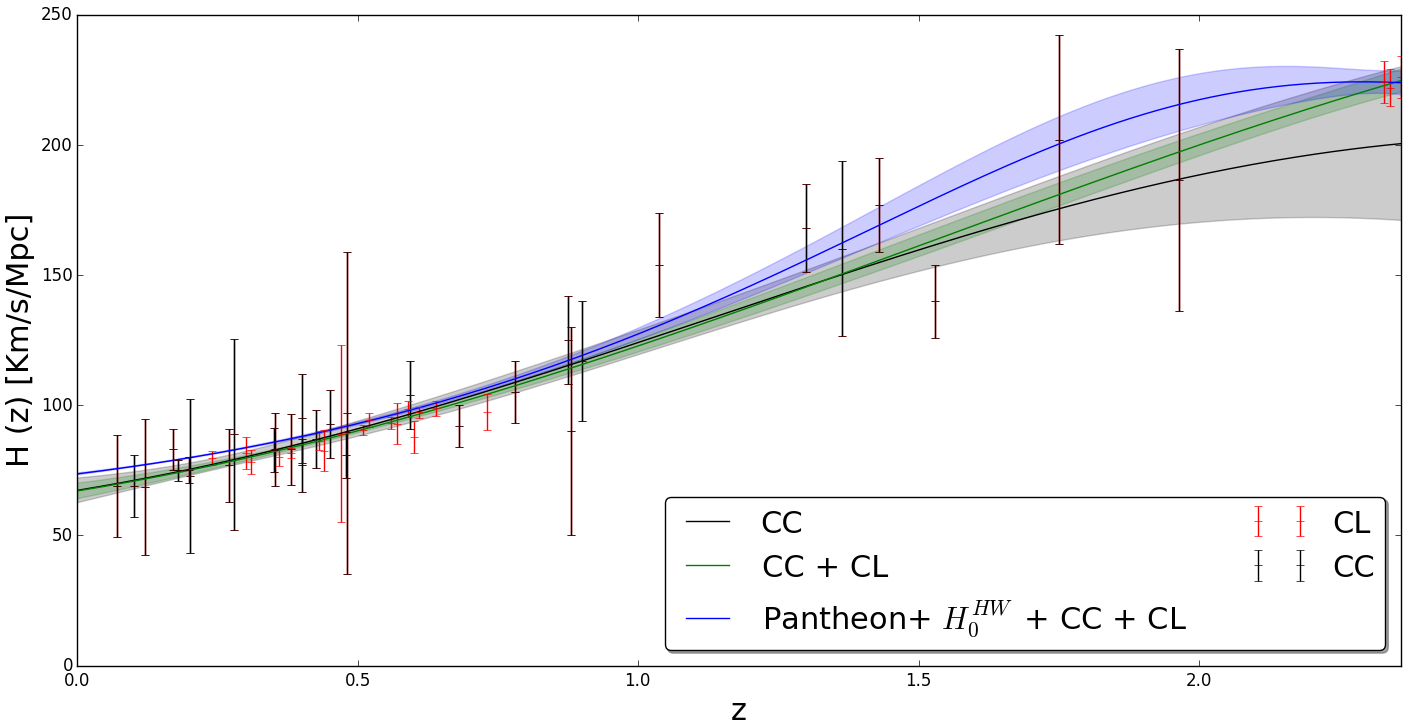} 
\includegraphics[scale=0.21]{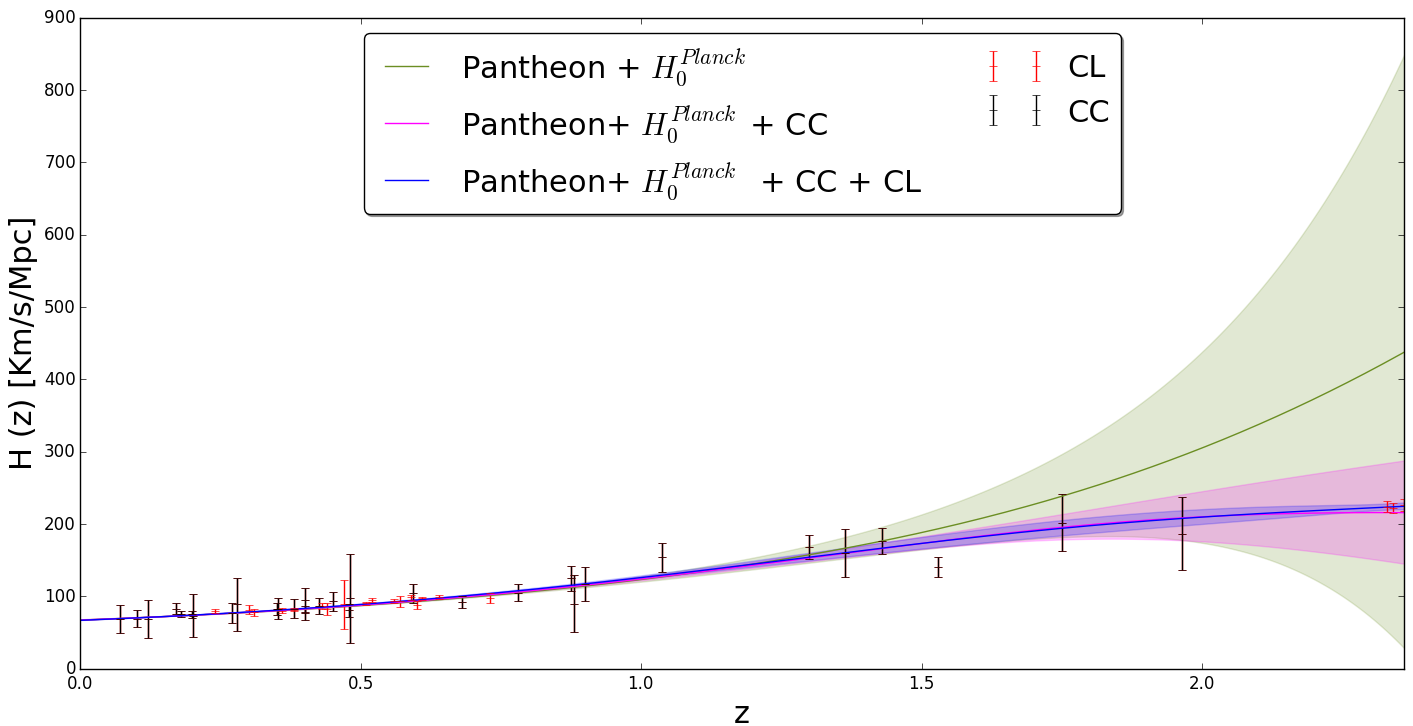} 
\includegraphics[scale=0.21]{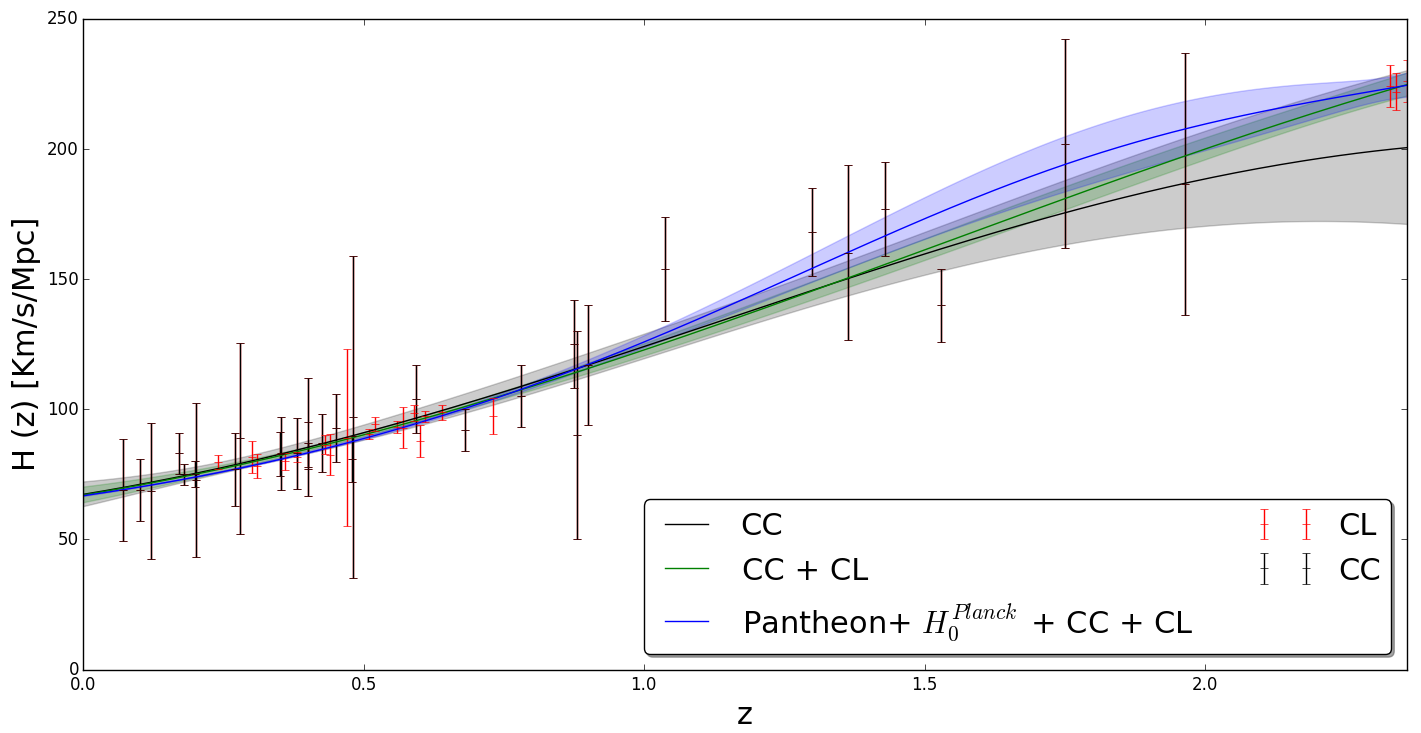} 
\includegraphics[scale=0.21]{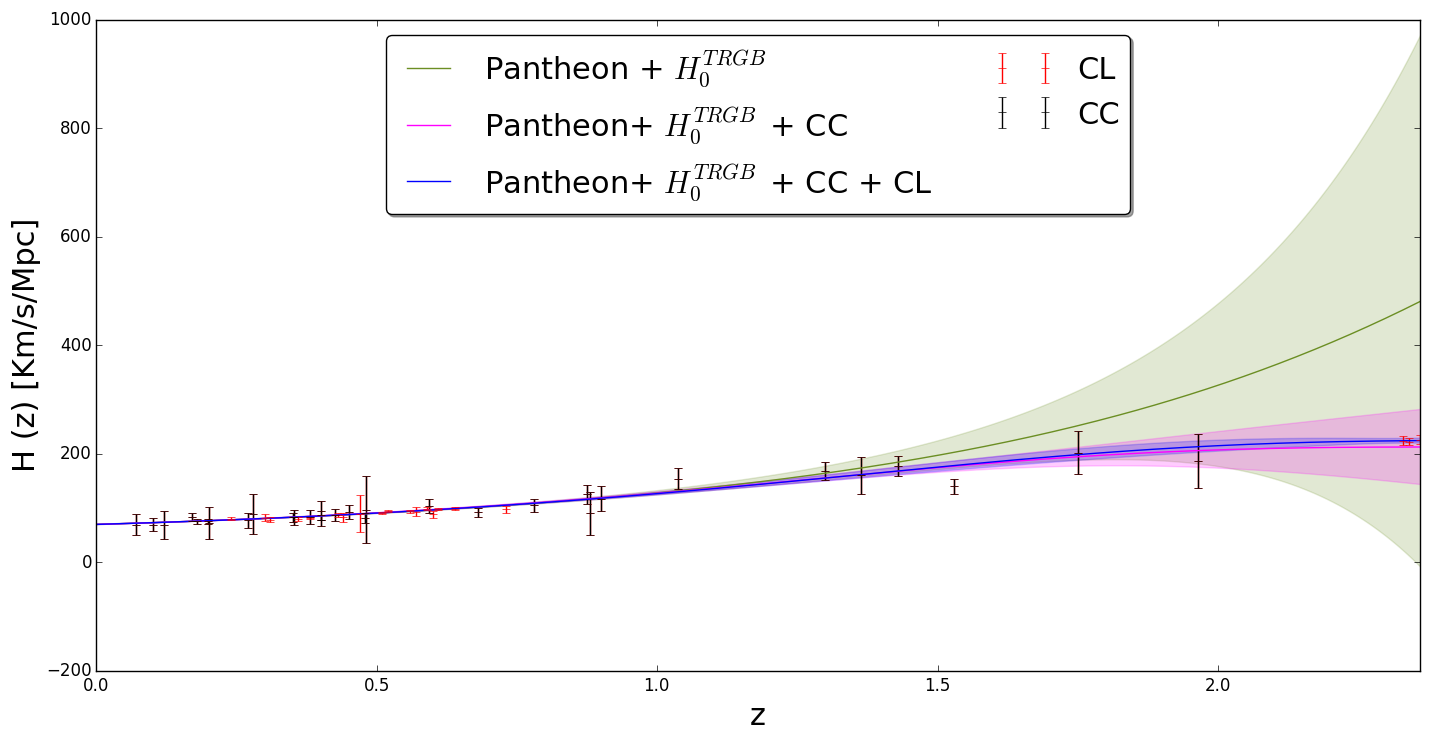} 
\includegraphics[scale=0.21]{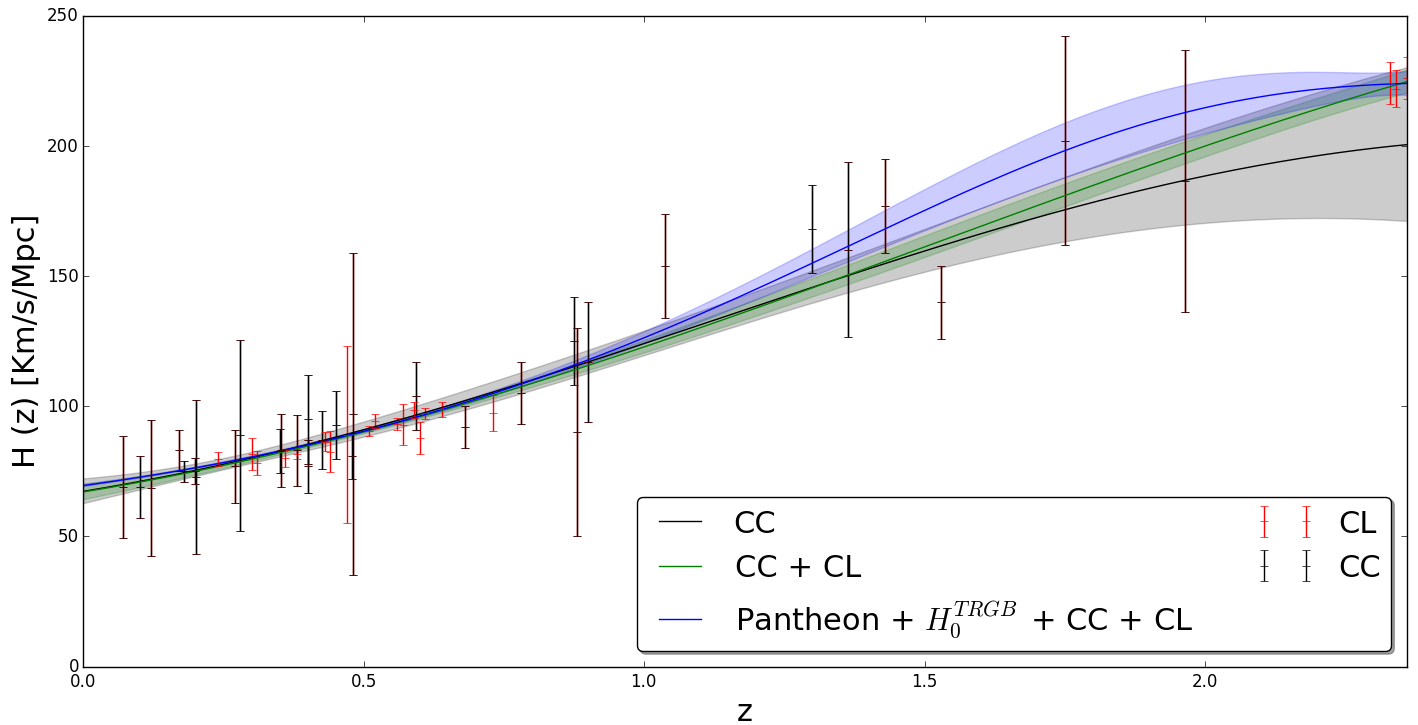} 
    \caption{Reconstructions of $H(z)$ with the squared exponential kernel (\ref{eq:SE}) and its corresponding uncertainties up to 1$\sigma$ with the four $ H_{0}$ prior. \textit{From top to bottom:} Using $H^{\text{R}}_{0}$, using $H_ {0}^{\text{HW}}$, using $H_ {0}^{\text{Planck}}$, and using  $H_{0}^{\text{TRGB}}$.
    We denote: (1) Pantheon+$H_{0}$ prior (green olive color), (2) Pantheon+$H_{0}$ prior+ CC (red color), and (3) Pantheon+$H_0$ prior + CC + CL (blue color). We include also the observational data given by the green olive color and red color points which describe CL and CC measurements, respectively.     }%
    \label{fig:example1}
\end{figure*}

To perform the reconstructions of $ H (z) $, it is necessary to transform the measurements. We start by  taking the  data of the Pantheon sample, once  is calibrated it gives us information about the distance module $ \mu (z) $, which is related to the luminous distance through the following equation
\begin{equation}
    d_{L}=  10^{(\mu(z)-25)/5},\label{eq:mu}
\end{equation}
we substitute the data of $\mu (z) $ in (\ref{eq:mu}), and  we obtain
\begin{equation}
        d_ {L} =\overline{d}_{L} + \sigma_{d_{L}},\label{eq:extr}
\end{equation}
where $\overline{d}_{L}$ is the mean value of $d_{L}$ and $\sigma_{d_{L}}$ is the  uncertainty  that results from the error propagation. In order to simplify the analysis, it is useful to define the following quantity
\begin{equation}
d_{p}\equiv \frac{d_{L} }{(1+z)c},
\label{eq:dpm}
\end{equation}
where $c$ is the speed of light. When we substitute (\ref{eq:extr}) in  (\ref{eq:dpm}) allows us to transform the data from $\mu (z) $ into information about the function $d_{p}(z)$. Moreover, if we assume that the universe is spatially flat and we make use  of  (\ref{eq:dpm}) we can show that
\begin{equation}
 d_{p}'=\frac{1}{H(z)},
 \label{eq:dppr}   
\end{equation}
which allows us to transform the measurements of CC and CL to information about $ d_ {p}'(z)$. To do so, we substitute the  measurements of $H(z)$ in (\ref{eq:dppr}) to obtain
\begin{equation}
    d'_{p}(z)=\frac{1}{\overline{H}(z) }\pm \frac{\sigma_{H}(z)}{\overline{H}^{2}(z)},\end{equation}
where  $\overline{H}(z)$ is the mean value of the observations of $H(z)$ with redshift $z$ and $\sigma_{H}(z)$ is its corresponding uncertainty at 1$\sigma$.  
\\\\
Once the process of transforming the measurements is performed we can notice that Pantheon data gives information about the function $ d_{p}(z)$ and the CC and CL measurements give information about $d'p(z)$. We can combine these data sets and apply a GP. With this process we can reconstruct the function    $d_p$ as a function of $z$ in the redshift range of the observable, e.g. for Pantheon compilation $0<z<2.36$,  using this function we can  get $ H(z) $ from 
\begin{equation}
    H(z)=\frac{1}{d'_{p}(z)}= \frac{1}{\overline{d'}_{p}(z) }\pm \frac{\sigma_{d'_{p}}(z)}{\overline{d'}^{2}_{p}(z)},\label{eq:8}
\end{equation}
where $\overline{d'_{p}}$  is the mean value of the derivative of the reconstruction and $ \sigma_ {d'_{p}}$ is the standard deviation of $ d'_{p}$.  In Figure \ref{fig:example1} we show the results for each one of the reconstructions using the kernels (\ref{eq:SE})-(\ref{eq:CF}) for three sets of data: (1) Pantheon+$H_0$ prior, (2) Pantheon+$H_0$ prior + CC, and (3) Pantheon+$H_0$ prior + CC + CL. 

Furthermore, since the reconstructions have been calibrated with different priors for $ H_ {0} $, we can compute the $\chi^2$-statistics for each case. Therefore, we proceed to define the following quantities:
\begin{align}
\chi^{2}_{\text{Pantheon}}= \sum_{i}^{N_{\text{Pantheon}} }\frac{    \left (  \mu(z_{i})_{\text{obs}}- \mu(z_{i})_{\text{recons}} \right )^{2}    }{\sigma(z_{i})_{\text{obs,Pantheon}}^{2}},\\
\chi^{2}_{\text{CC}}= \sum_{i}^{N_{\text{CC}}}\frac{    \left ( H(z_{i})_{\text{obs}}- H(z_{i})_{\text{recons}} \right )^{2}    }{\sigma(z_{i})_{\text{obs,CL}}^{2}},\\
\chi^{2}_{\text{CL}}= \sum_{i}^{N_{\text{CL}}}\frac{    \left ( H(z_{i})_{\text{obs}}- H(z_{i})_{\text{recons}} \right )^{2}    }{\sigma(z_{i})_{\text{obs,CL}}^{2}},\\
\chi^{2}_{H_{0}}= \frac{    \left ( H_{0,\text{obs}}- H_{0,\text{recons}} \right )^{2}    }{\sigma(z_{i})_{\text{obs},H_{0}}^{2}},
\end{align}
where $N$ is the number of  $D$ data point, $D_{\text{obs}}(z_{i})$ is the data point observed with redshift $z_{i}$ , $\sigma_{i}$ is the uncertainty associated with each measurement,  $D_{\text{recons}}(z_{i})$ is the value predicted by the reconstruction, and we define the following quantity 
\begin{equation}
    \chi^{2}_{\text{Total}}=\chi^{2}_{\text{Pantheon}} +\chi^{2}_{\text{CC}}+\chi^{2}_{\text{CL}}+\chi^{2}_{H_{0}}. \label{eq:chitotal}
\end{equation}

%%%%%%%%%%%%%%%%%%%%%%%%%%%%%%%%%%%%%%%%%%%%
%%%%%%%%%FIGURES RECONSTRUCTIONS%%%%%%%%%%%%
%%%%%%%%%%%%%%%%%%%%%%%%%%%%%%%%%%%%%%%%%%%%

\begin{figure*}
    \centering
\includegraphics[scale=0.21]{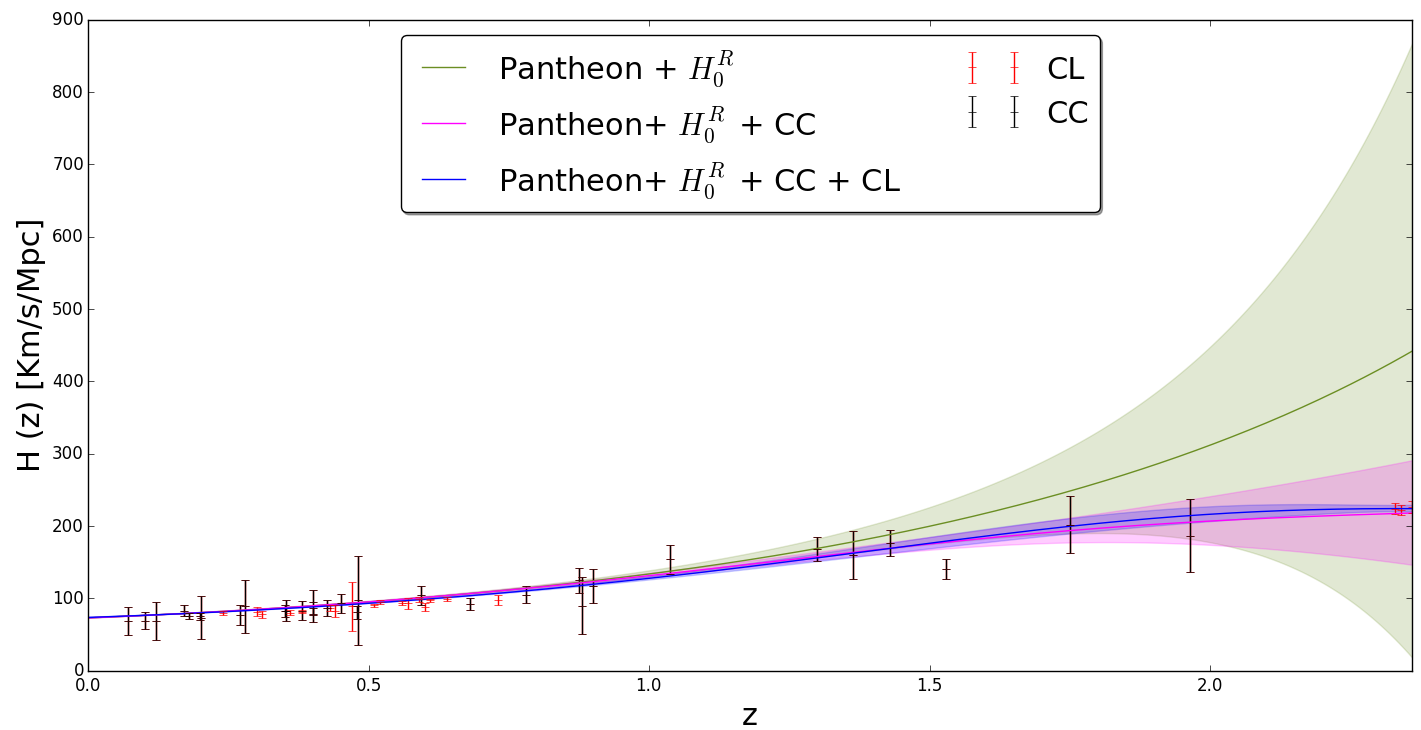} 
\includegraphics[scale=0.21]{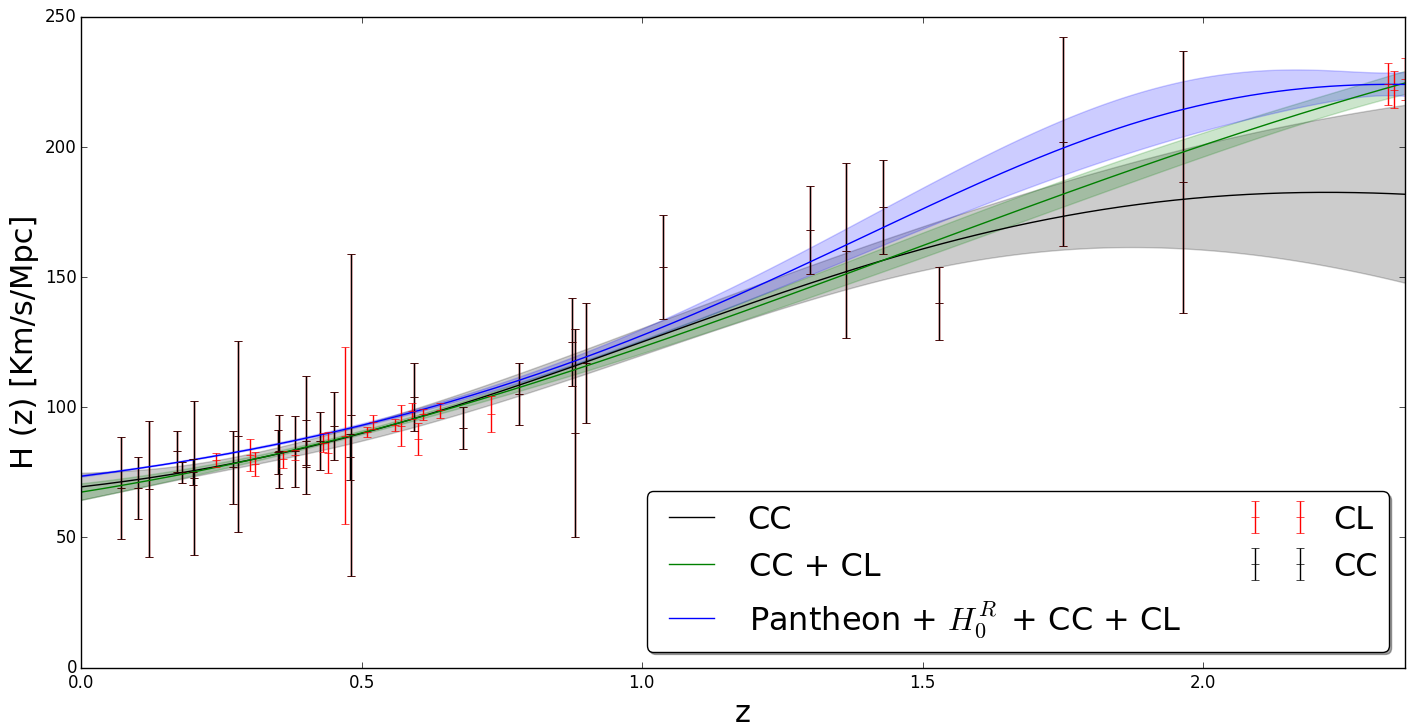} 
\includegraphics[scale=0.21]{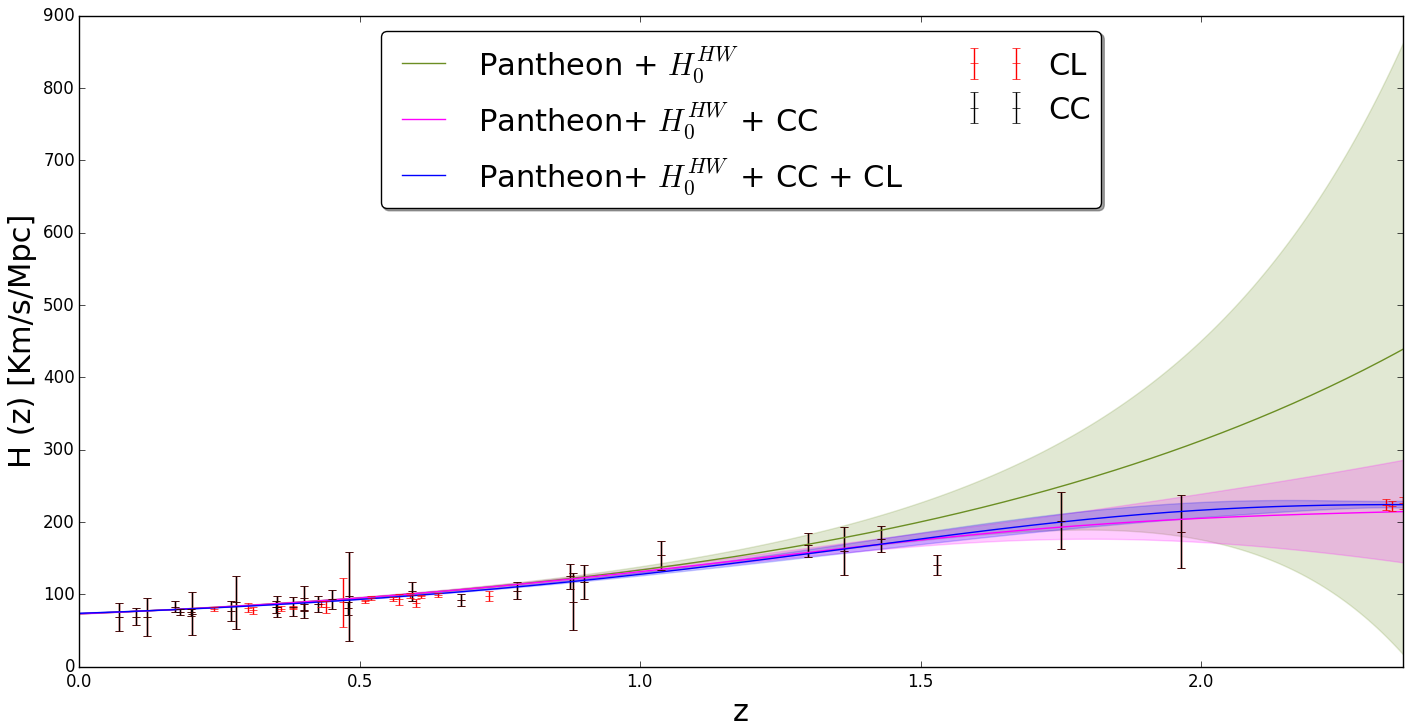} 
\includegraphics[scale=0.21]{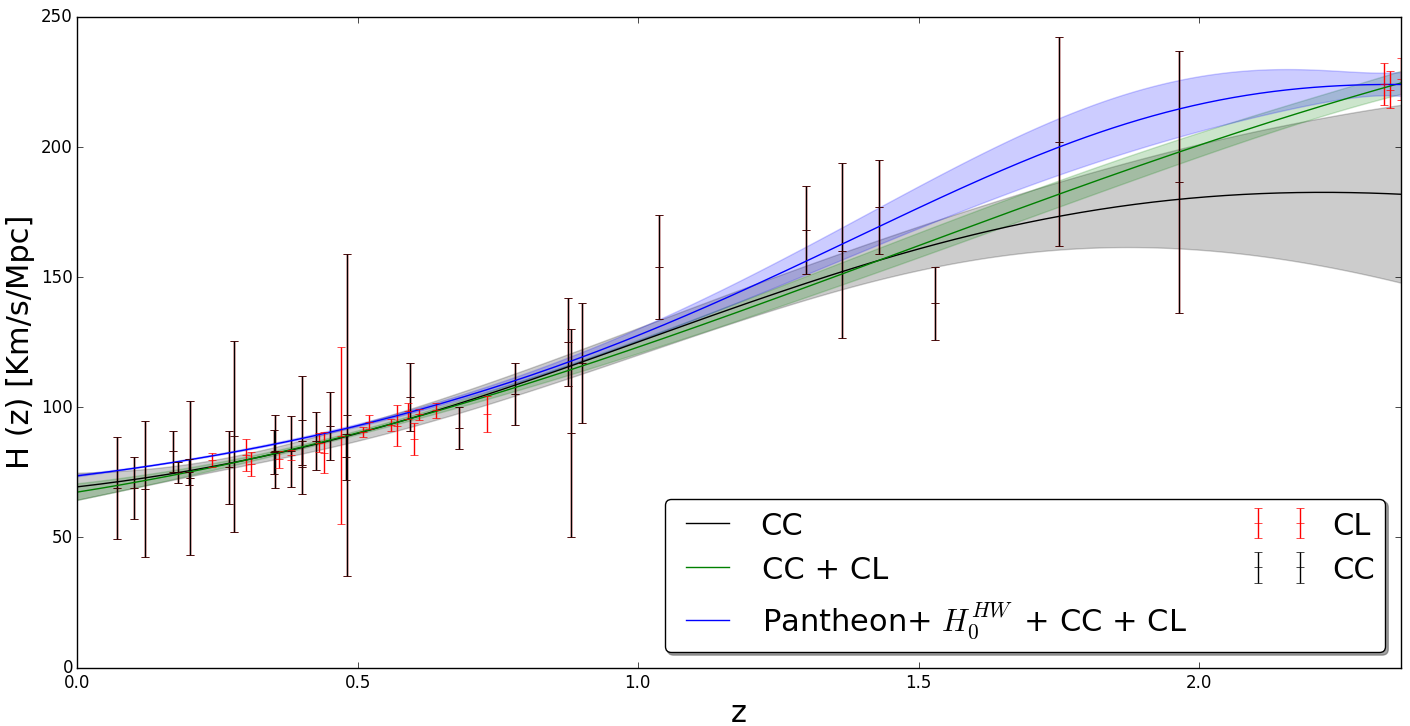} 
\includegraphics[scale=0.21]{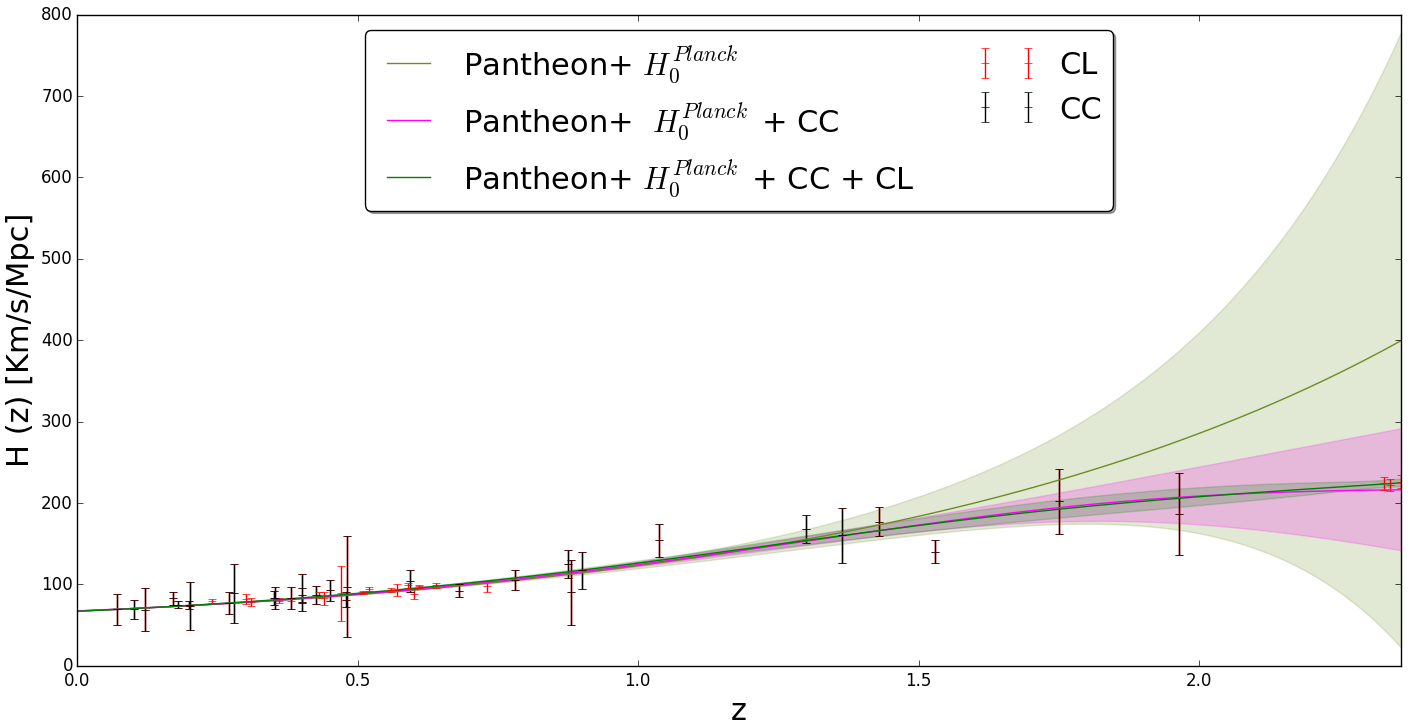}
\includegraphics[scale=0.21]{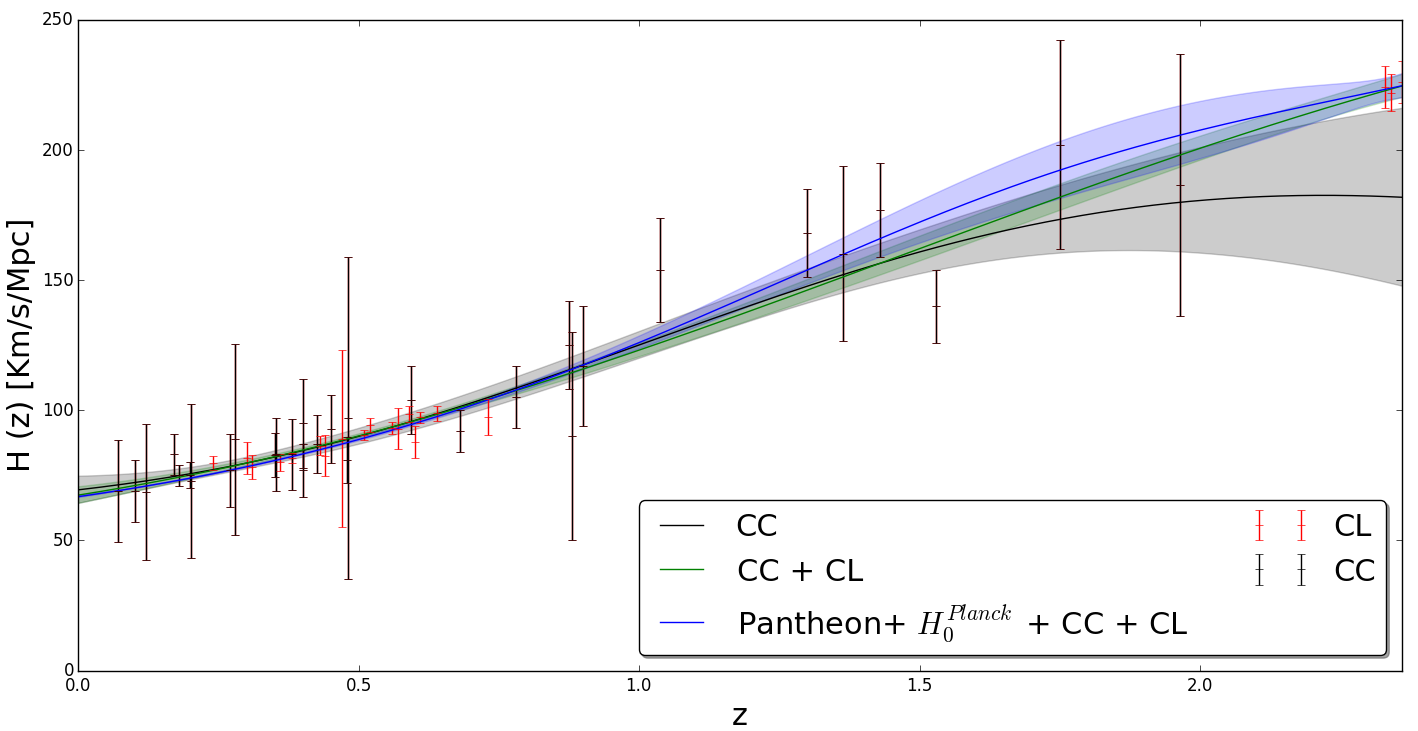} 
\includegraphics[scale=0.21]{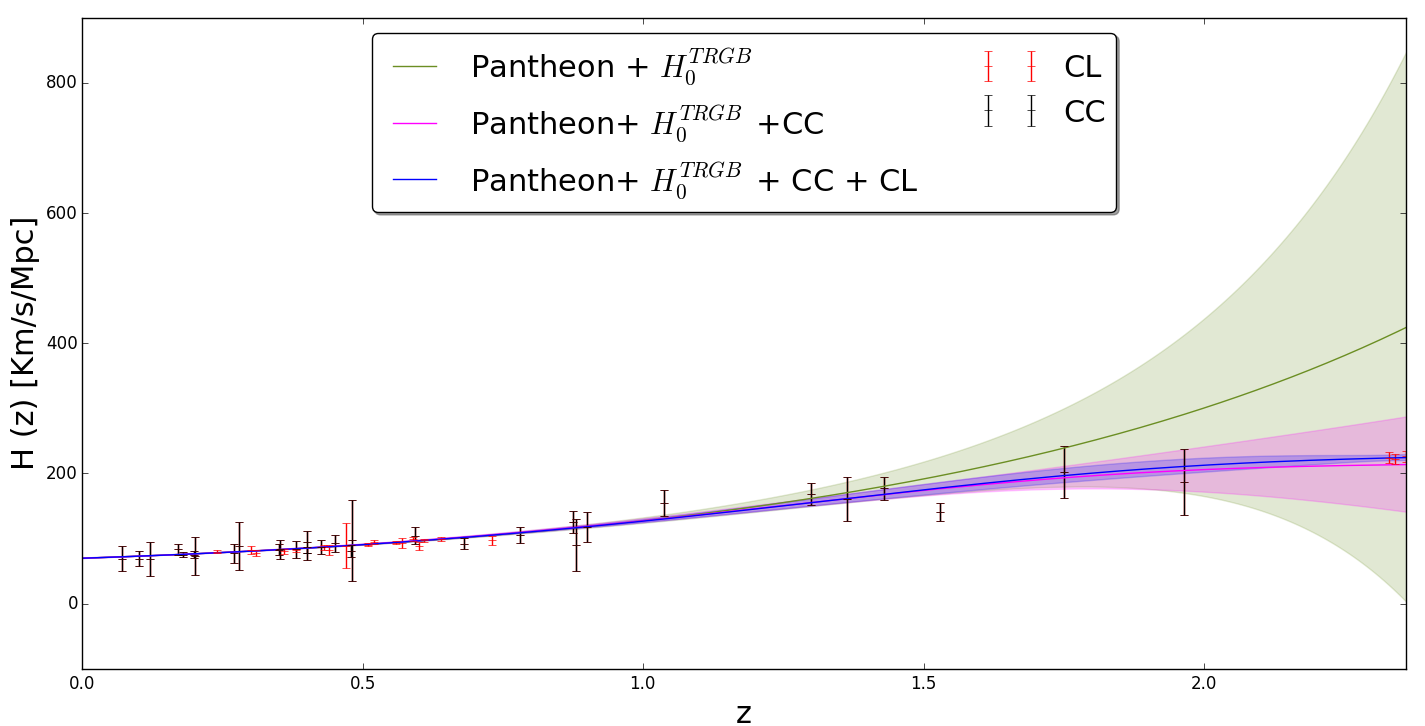} 
\includegraphics[scale=0.21]{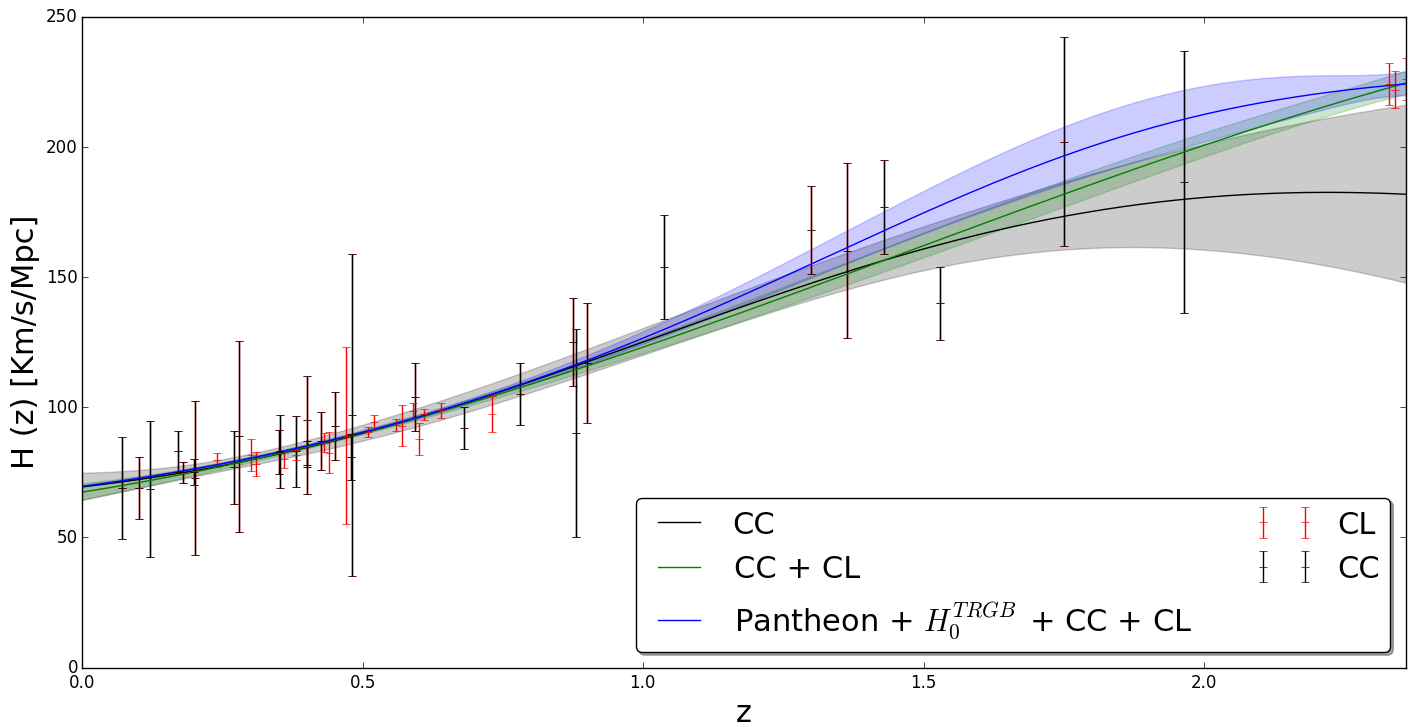} 
    \caption{ Reconstructions of $H(z)$ with the Cauchy function kernel (\ref{eq:CF}) and its corresponding uncertainties up to 1$\sigma$ with the four $ H_{0}$ prior. \textit{From top to bottom:} Using $H^{\text{R}}_{0}$, using $H_ {0}^{\text{HW}}$, using $H_ {0}^{\text{Planck}}$, and using  $H_{0}^{\text{TRGB}}$.
    We denote: (1) Pantheon+$H_{0}$ prior (green olive color), (2) Pantheon+$H_{0}$ prior+ CC (red color), and (3) Pantheon+$H_0$ prior + CC + CL (blue color). We include also the observational data given by the green olive color and red color points which describe CL and CC measurements, respectively. 
}
    \label{fig:example5}
\end{figure*}

In Table \ref{table:3} we show the results for $\chi^{2}_{\text{Total}}$ (\ref{eq:chitotal}) for each reconstruction case. Notice that the reconstruction that better describes the data sets is the one calibrated with $H_{0}^{\text{TRGB}}$, therefore this reconstruction can guide us in the search for a model that could solve the $H_{0}$ tension with the observations of the late-time universe.

\begin{table}
\centering
\begin{tabular}{  |p{5.5cm} |p{5.5cm} |p{3.3Cm} |}
\hline
\multicolumn{3}{|c|}{ $ \chi^{2}_{\text{Total}}$(\ref{eq:chitotal})  for the different reconstructions.} \\
\hline
Data sets    &  Squared exponential kernel (\ref{eq:SE})  &Cauchy kernel (\ref{eq:CF})  \\
\hline
Pantheon+ $H_{0}^{\text{R}}$ + CC+ CL    &  1151.88      &  1150.53 \\
Pantheon+ $H_{0}^{\text{HW}}$ + CC+ CL    &   1065.38   &  1063.97    \\
Pantheon+ $H_{0}^{\text{Planck}}$ + CC+ CL   &   1138.33 &  1137.24 \\
Pantheon + $H_{0}^{\text{TRGB}}$ + CC+ CL  &    952.90   & 952.04  \\
\hline
\end{tabular}
\caption{$\chi^2$-statistics for the full data sets reconstructed  Pantheon+ $H_{0}$ + CC+ CL (first column) using the $H_0$ priors for the squared exponential kernel (second column) and the Cauchy kernel (third column), using (\ref{eq:chitotal}). }
\label{table:3}
\end{table}

%%%%%%%%%%%%%%%%%%%%%%%%%%%%%%%%%%%%%%%%%%%%
%%%%%%%%%%%%%%%%%%%%%%%%%%%%%%%%%%%%%%%%%%%%

\subsection{GP calibrated reconstruction methods and comparison tests}
To compare our model-independent reconstruction with other proposals in this direction, several methods have been presented in the literature to obtain reconstructions of the Hubble parameter \cite{gomez2018h0,Briffa:2020qli,mehrabi2020does,Escamilla-Rivera:2015odt}. Moreover, some of them present some disadvantages with respect to our methodology:

\begin{enumerate}
\item  \textbf{SNIa/QSO cosmokinetic reconstruction method} \cite{mehrabi2020does}. In this proposal a GP reconstruction is based to cover a large redshift range by using quasars \cite{risaliti2019cosmological,risaliti2015hubble} calibrated measurements with Gamma Ray Bursts \cite{demianski2017cosmology,demianski2017cosmology,amati2013measuring}. In comparison to our proposal in the SNIa/QSO cosmokinetic reconstruction method is not indicated the value of $H_{0}$ used to calibrate Pantheon data, therefore, a straightforward bias emerges as a consequence of this prior.
    \item  \textbf{Covariance functions reconstruction method} \cite{Briffa:2020qli}. In this research the main goal was to obtain constraints on Teleparallel Gravity models \cite{Escamilla-Rivera:2019ulu} through GP using the CC data set, BAO data \cite{alam2017clustering,des2017baryon,zhao2019clustering}, and the Pantheon compressed data set \cite{riess2018type} including 15 SNeIa measured by CANDELS and CLASH Multy-Cycle Treasury program (hereafter, PMCT). This data set compresses all the information of SNeIa  in only six data points for $E(z) = H (z) / H_ {0}$. However, in  this method it was used  only five points, since the last one  is non-Gaussian distributed. 
 Since the PMCT data is given in terms of $ E (z) $, the method requires the calibration of this sample before being reconstructed. To calibrate the PCMT data, first they  perform a GP on the CC data, this allows them to get a value for $ H_ {0} $. Using this value of $ H_ {0} $ they calibrate the PCMT data.  Second, it is possible to perform a GP combining CC+PMCT data to obtain a new $H_{0}$ value and afterwards they use this value to re-calibrate the PMCT sample. Finally, the process is  repeat it  until the obtained value of $H_ {0}$ converges. 
It is important to notice that this method found that $\Lambda$CDM is within the 1$\sigma$ region for all the reconstructions, regardless of the data sets used in their analysis. 
    \item  \textbf{Reconstruction with weighted polynomial regression method} \cite{gomez2018h0}. In this proposal was presented a method to  
obtain constraints on the late-cosmic expansion. Based on CC+PMCT data samples and under the consideration to circumvent the use of $H(z)$ measurements obtained from BAO data. It is possible to calibrate the PMCT data and use the same procedure as in the GP covariance functions reconstruction method. The results of this method suggest that the reconstructions  have  a preference for low values of $H_{0}$.
\end{enumerate}
The methods discussed in (2) and (3)  have two main disadvantages with respect to our proposal:
\begin{itemize}
    \item   The number of data points used in methods (2) and (3) to perform the reconstructions is below 50 points, which is much lower than the number of points used in our analysis, which correspond to 31 points for the  CC data set, 20 points for CL, 1048 for Pantheon and one prior for calibration. The low number of points used in their reconstruction results in a 1$\sigma $ uncertainty for their reconstructions significantly bigger than ours, for example for every reconstruction performed with the full data set our uncertainty in $ H_ {0} $ is reduced by at least a factor of 2 with respect to these methods.
   \item The iterative process carried out to calibrate the PMCT data causes a preference for small values of $H_{0}$, to see this let us define  the following \cite{camarena2018impact} quantity 
\begin{equation}
    T\equiv\frac{| H_{0i}-H_{0j} |}{\sqrt{\sigma^{2}_{i}+ \sigma^{2}_{j} }},
\end{equation}
where $H_{0i}$ and $H_{0j}$ are the mean values of the Hubble parameter and $\sigma_{i}$, $\sigma_{j}$ are the corresponding 1$\sigma$ uncertainties.  By considering $ H_ {0}^{\text{Planck}}$, $ H_ {0}^{\text{R}} $ and $ H_ {0}^{\text{Prior}} = 69.39 \pm 5.18 $\footnote{We are using the $H_{0}$ for the Cauchy kernel obtained form Cosmic Chronometers obtained in \cite{gomez2018h0}.}, we calculate  the tension  between them:
\begin{itemize}
\item Tension between $H_{0}^{\text{Planck}}$ and $H_{0}^{\text{Prior}}$:  $T=0.382$.
  \item Tension between $H_{0}^{ \text{R}}$ and  $H_{0}^{\text{Prior}}$:  $T=0.76$.
\end{itemize}
We can see that the prior  gives a priority to $ H_{0}^{\text{Planck}}$ (low values of $H_{0}$). By repeating the process several times this preference increases and seems to be confirmed by their final results reported. We avoid this problem in our proposal by calibrating Pantheon data for every choice of $H_{0}$ that  we consider. This is extremely important for our work since we are studying the Hubble tension effects.
\end{itemize}

In Tables \ref{table:1}-\ref{table:2} we report our results for $H_{0}$, along with a comparison with three methods described. The results are present as follows:  In the first column of Table \ref{table:1} we show the data sets used to perform our reconstructions, in the second column we show the value of $H_{0}$ obtained from each of the reconstructions using the squared exponential kernel, and in columns 3, 4 and 5, we present a comparison with methods mentioned. Table \ref{table:2} follows has the same structure as Table \ref{table:1}, the only difference is that the Table \ref{table:2} presents the results using the Cauchy kernel, finally the symbols appearing in the tables denotes the following:

\begin{itemize}
    \item[] $\bigstar$: denotes that our result  agrees with the Reference reported at the top of the Table.
    \item[] $\bigotimes$: denotes that our result  does not match the Reference at the top of the Table.
        \item[] $-$: denotes  that our  result was not reported in the Reference at the top of the Table.
    \item[] $\bullet$: denotes that our result has not been previously reported in the literature.
\end{itemize}

\bigskip 
\begin{table}
\centering
\begin{tabular}{  |p{5.4cm} |p{2.4cm} |p{1.9cm}|p{2.4cm} |p{1.9cm}|}
\hline
\multicolumn{5}{|c|}{$H_{0}$ results for the squared exponential covariance function (\ref{eq:SE}) } \\
\hline
Data sets     & Our method   &  Method (1) \cite{mehrabi2020does} &  Method (2) ~ \cite{Briffa:2020qli}  &   Method (3) \cite{gomez2018h0} \\
\hline
CC    & $ 67.32 \pm 4.74 $     & --   &    $\bigstar$ & $\bigstar$ \\
CC + CL    & $67.08 \pm 3.01 $ $\bullet$& --   &    -- & -- \\
\hline
Pantheon+ $H_{0}^{R}$  & $72.94 \pm 0.40 $  & -- &    -- &   -- \\
Pantheon+ $H_{0}^{R}$ + CC    &  $  72.83 \pm  0.41$    & --  &  $\bigstar$ & --\\
Pantheon+ $H_{0}^{R}$ + CC + CL    &  $73.48 \pm 0.36 $  $\bullet$   & --    &    -- & -- \\
\hline
Pantheon+ $H_{0}^{\text{HW}}$  & $ 73.00 \pm 0.41$ $\bullet$  & -- &   -- &   -- \\
Pantheon+ $H_{0}^{\text{HW}}$ + CC  &  $  72.97 \pm  0.42$    & --  &  $70.85 \pm 1.19$ $\bigotimes$& -- \\
Pantheon+ $H_{0}^{\text{HW}}$ + CC + CL    &  $ 73.63 \pm  0.38 $  $\bullet$   & --    &    -- & -- \\
\hline
Pantheon+ $H_{0}^{\text{Planck}}$  &  $  67.11\pm 0.37  $  $\bullet$ & -- &   -- &   -- \\
Pantheon+ $H_{0}^{\text{Planck}}$ + CC    &  $ 67.10 \pm 0.38 $  $\bullet$  & --   & -- & --\\
Pantheon+ $H_{0}^{\text{Planck}}$ + CC + CL &  $  66.77\pm 0.33 $  $\bullet$   & --    &    -- & -- \\
\hline
Pantheon+ $H_{0}^{\text{TRGB}}$  & $ 69.53 \pm 0.41 $  $\bullet$ & -- &   -- &   -- \\
Pantheon+ $H_{0}^{\text{TRGB}}$ + CC    &  $  69.51 \pm  0.42$    & --  & $\bigstar$  & -- \\
Pantheon+ $H_{0}^{\text{TRGB}}$ + CC + CL &  $69.53 \pm 0.37 $  $\bullet$   & --    &    -- & -- \\
\hline
\end{tabular}
\caption{Values of $H_{0}$ [km/s/Mpc] with their corresponding 1$\sigma$ uncertainties inferred via the GP reconstructions. Notice that our result for the Pantheon+ $H_{0}^{\text{HW}}$+ CC data set is at $1.68\sigma$ tension with respect method (3) \cite{Briffa:2020qli}. }
\label{table:1}
\end{table}

\begin{table}
\centering
\begin{tabular}{  |p{5.4cm} |p{2.4cm} |p{1.9cm}|p{2.4cm} |p{1.9cm}|}
\hline
\multicolumn{5}{|c|}{$H_{0}$ results for the Cauchy covariance function (\ref{eq:CF}) } \\
\hline
Data sets     & Our method  & Method (1) \cite{mehrabi2020does} & Method (2) ~\cite{Briffa:2020qli}  &   Method (3) \cite{gomez2018h0} \\
\hline
CC    & $ 69.43 \pm 5.23 $     & --   &    $\bigstar$ & $\bigstar$ \\
CC + CL    & $67.41 \pm 3.17 $ $\bullet$    & --   &    -- & -- \\
\hline
Pantheon+ $H_{0}^{R}$  & $ 72.92 \pm 0.42 $  & -- &    -- &   -- \\
Pantheon+ $H_{0}^{R}$ + CC    &  $ 72.86 \pm 0.42 $    &  --  &   $\bigstar$ & --\\
Pantheon+ $H_{0}^{R}$ + CC + CL    &  $73.51 \pm 0.37 $  $\bullet$  & --  &   -- & --\\
\hline
Pantheon+ $H_{0}^{\text{HW}}$  & $ 72.98 \pm 0.43 $ $\bullet $  & -- &    -- &   -- \\
Pantheon+ $H_{0}^{\text{HW}}$ + CC    &  $  73.00 \pm 0.44 $     &  --   &  $70.89 \pm 1.20$ $\bigotimes$ & --\\
Pantheon+ $H_{0}^{\text{HW}}$ + CC + CL    &  $ 73.67 \pm  0.39$  $\bullet$  & --  &   -- & --\\
\hline
Pantheon+ $H_{0}^{\text{Planck}}$  & $ 67.10\pm 0.39  $ $\bullet$  & -- &    -- &   -- \\
Pantheon+ $H_{0}^{\text{Planck}}$ + CC    &  $ 67.13 \pm  0.39$ $\bullet$    & --  &   -- & --\\
Pantheon + $H_{0}^{\text{Planck}}$ + CC + CL    &  $ 66.81 \pm 0.35 $  $\bullet$  & -  &   -- & --\\
\hline
Pantheon+ $H_{0}^{\text{TRGB}}$  & $ 69.52 \pm 0.43 $ $\bullet$  & -- &    --  &   --\\
Pantheon+ $H_{0}^{\text{TRGB}}$ + CC    &  $ 69.55 \pm 0.44 $    &  --   &    $\bigstar$ &-- \\
Pantheon+ $H_{0}^{\text{TRGB}}$ + CC + CL    &  $69.58 \pm 0.38 $  $\bullet$  & --  &   -- & --\\
\hline
\end{tabular}
\caption{Values of $H_{0}$ [km/s/Mpc] with their corresponding 1$\sigma$ uncertainties inferred via the GP reconstructions. Notice that our result for the Pantheon+ $H_{0}^{\text{HW}}$+ CC data set is at  1.65$\sigma$ tension with respect method (3) \cite{Briffa:2020qli}. }
\label{table:2}
\end{table}

%%%%%%%%%%%%%%%%%%%%%%%%%%%%%%%%%%%%%%%%%%%%
%%%%%%%%%%%%%%%%%%%%%%%%%%%%%%%%%%%%%%%%%%%%
\section{Model-independent constraints on density parameters}
\label{sec:model_constraints}

Since the  Hubble constant and the dark energy density at present time are 
correlated  \cite{Aghanim:2018eyx},  it is  expected that the Hubble tension  also manifests itself as a tension between the value of the dark energy density   inferred from the fit of $\Lambda$CDM to the observations of the  CMB and the value obtained with the reconstructions. 
To obtain the dark energy density from the reconstructions we will assume the following: a spatially flat universe with matter content  modeled as a dust-like fluid, in addition  
we will  assume that  dark energy can  be modeled by an effective fluid  with  a parametrised equation of state $\omega(z)$. Under these assumptions the standard Friedmann equations have the following form
\begin{eqnarray}
&&H^{2}=H_{0}^2 [\Omega_{m}^{0}(1+z)^3 +\Omega_{\text{eff}}^{0}f(z) ]\label{eq:ome},\\
&&\frac{\ddot{a}}{a}=-\frac{H^{2}}{2}\left (\Omega_{m}+\Omega_{\text{eff}}(1+3\omega)  \right ),\label{eq:ac}  \\
&&    \dot{\rho}+3H(1+P)=0,
\end{eqnarray}
where the index ``0'' denotes quantities evaluated at present cosmic time.  Following the standard definition, $ \Omega_{\text{eff}} $  denotes the dark energy density parameter and  it changes as function of redshift as  $\Omega_{\text{eff}}(z) = \Omega_{\text{eff}}^{0}f(z)$, with
$f(z)= \exp \left (  3\int_{0}^{z} \frac{1+\omega}{1+z'}dz'\right )$. In particular for $ \Lambda$CDM $ f(z)=1 $ for all z. According to the constraint $ \Omega_{m} + \Omega_{\text{eff}} = 1 $, we can rewrite (\ref{eq:ome}) as 
 \begin{equation}
     H^{2}=H^{2}_{0}[(1-\Omega_{\text{eff}}^{0})(1+z)^3 +\Omega_{\text{eff}}^{0}f(z) ]\label{eq:H17}.
 \end{equation}   
If we evaluate (\ref{eq:H17}) at $ z = 0 $, we have that the function $f(z)$  must obey  $ f(z=0)=1$. With this ansatz we can rewrite (\ref{eq:ome}) as follows 
\begin{equation}
    H^{2}-H^{2}_{0}(1+z)^{3}=H_{0}^{2}\Omega_{\text{eff}}^{0}[f(z)-(1+z)^{3}]. \label{eq:nad}
\end{equation}
Notice that if $z=0$, both sides of the above equation vanish and therefore we cannot get any information from $ \Omega_{\text{eff}}^{0} $. However, we expect that $ H (z) $ is a smooth (differentiable) function, then $ f (z) $ does not have abrupt changes, i.e. $ f(z\approx 0) \approx 1 $. 
Moreover, in the limit $ z \rightarrow 0 $ we found that
\begin{equation}
    H^{2}-H^{2}_{0}(1+z)^{3}\approx H_{0}^{2}\Omega_{\text{eff}}^{0}[1-(1+z)^{3}]\label{eq:desco},
\end{equation}
therefore
\begin{equation}
    \Omega_{\text{eff}}^{0}\approx\frac{H^{2}-H^{2}_{0}(1+z)^{3}}{ H_{0}^{2}[1-(1+z)^3]} \label{eq:mnn1}.
\end{equation}
We proceed to define   $\Delta z$  as
\begin{equation}
\Delta z= \frac{z_{f} - z_{i}}{N},
\end{equation}
where $N$  is the number of points placed  in the reconstruction, and the subscripts $f$, $i$ denote final and initial values of $z$, respectively. Taking $z_{i}=0$, we see that  the smallest $z$  that we have access with the reconstruction is $\Delta z$, therefore
\begin{equation}
    \Omega_{\text{eff}}^{0}=\frac{H^{2}-H^{2}_{0}(1+\Delta z)^{3}}{ H_{0}^{2}[1-(1+\Delta z)^3]}.\label{eq:mnn2}
\end{equation}
We can obtain the error propagation of the latter equation by writing $ H_ {0} $ and $ H (\Delta z) $ as
\begin{eqnarray}
H_{0}&=&\overline{H}_{0}\pm \sigma_{H_{0}},\\
H(\Delta z)&=&\overline{H}_{1}\pm \sigma_{H_{1}},
\end{eqnarray}
where the first term (denoted by a bar on top it) represents the mean value ,and the second is the 1$\sigma $ uncertainty of $ H_ {0} $ and $ H (\Delta z) $, respectively. With these definitions we can perform the error propagation to obtain
\begin{equation}
\Omega_{\text{effe}}^{0}\approx \frac{\sqrt{2}}{(1+\Delta z)^{3}-1} \left ( \frac {  \overline{H}_{1}^{2}} {\overline{H}_{0}^{4}}\sigma_{H_{1}}^{2} + \frac {
 \overline{H}_{1}^{4}} {\overline{H}_{0}^{6}}\sigma_{H_{0}}^{2} \right )^{1/2}, 
\end{equation}
by definition $\Omega_{\text{effe}}^{0}$ is the $1\sigma$ error for $\Omega_{\text{eff}}^{0}$. We can rewrite this error in terms of the number of points $N$ in the reconstruction
\begin{equation}
\Omega_{\text{ effe}}^{0}\approx \frac{\sqrt{2} N^{3}}{(N+ z_{f})^{3}-N^{3}} \left ( \frac {  \overline{H}_{1}^{2} } {\overline{H}_{0}^{4}}\sigma_{H_{1}}^{2} + \frac { \overline{H}_{1}^{4} } {\overline{H}_{0}^{6}}\sigma_{H_{0}}^{2} \right )^{1/2},\label{eq:na}
\end{equation}
from where it is clear that the error depends on the number $ N $ of points  placed in the reconstruction, which leads us to the following issue: \textit{we cannot increase the number of data points in the reconstruction arbitrarily since this would increase the statistical uncertainty in the calculation of $ \Omega_ {\text{eff}} ^ {0} $, but either we can  make $ N $ too small or (\ref{eq:mnn1}) would not remain valid}.
\begin{figure}
    \centering
    \includegraphics[scale=0.7]{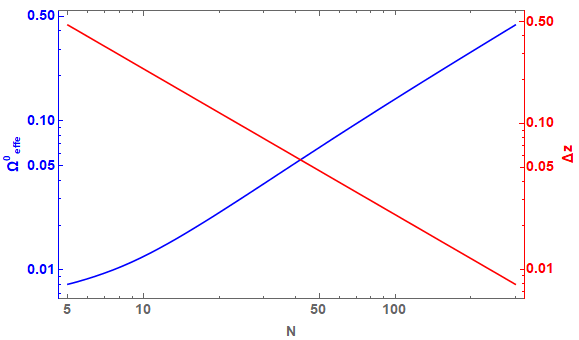}
    \caption{In blue it is shown how $\Omega_ {\text{eff}}^{0} $ varies as a function of the number of points $ N $ placed in the reconstruction, while in red it is shown how $ \Delta z$ varies as function of $N$. The intersection indicates which is the $ N $ that minimizes uncertainty and at the same time gives us a $\Delta z $ small enough to have a decent approximation. We get  that $ N_ {\text{optimal}} \equiv N_{0} \approx 43 $, which corresponds to $ \Delta z = 2.37/N_{0}. $ }
    \label{fig:my_labelmm}
\end{figure}
In order to solve this issue we define the following quantity:
\begin{equation}
    E\equiv\frac{H_{\text{approx}}(\Delta z)}{H_{\text{recons}}(\Delta z)} \label{eq:estimador},
\end{equation}
where $ H_{\text{approx}} $ is the $ H (z) $ given by  (\ref{eq:desco}) and $ H_{\text{recons}} $ is the $ H (z) $ given by the reconstruction. If $ E = 1 $,   (\ref{eq:mnn1}) is completely valid. If $ E = 1 $, (\ref{eq:mnn1}) is completely valid, if $ E <1 $ we will underestimate the value of $ \Omega_{\text{effe}}^ {0} $ and if 
$E>1$ we will overestimate the value $ \Omega_{\text{effe}}^ {0} $, since the Eq.~(\ref{eq:mnn1}) is no longer valid. Therefore, $ E $ gives us a measure of \textit{how good} the  approximation  given by  (\ref{eq:mnn1}) is. We present the results for $ \Omega_{\text{eff}} $ in  the following form:
\begin{equation}
    \Omega_{\text{eff}}^{0}=\overline{\Omega}_{\text{eff}}^{0}\pm \Omega_{\text{effs}}^{0}\pm \Omega_{\text{effn}}^{0},\label{eq:cal}\end{equation}
where $ \overline {\Omega}_{\text{eff}}^{0} $ is the mean value of $ \Omega_ {\text{eff}}^{0} $, $ \Omega_ {\text{effs}} ^ {0} $ is the uncertainty at 1$\sigma $ and $ \Omega_ {\text{effn}}^{0} $ is the error due to the approximation given by (\ref{eq:mnn1}). This error is  obtained through  (\ref{eq:estimador}). In Table \ref{table:31} we show the results of calculating (\ref{eq:cal}) with the reconstructions. It is interesting to note that $\Omega_{\text{effn}}^{0}<<\Omega_{\text{effs}}^{0}$ regardless of the reconstruction.
\begin{table}
\centering
\begin{tabular}{  |p{5.5cm} |p{5.5cm} |p{4cm} |}
\hline
\multicolumn{3}{|c|}{ Constrains on $\Omega_{\text{eff}}^{0}$ via Gaussian Processes } \\
\hline
Data sets used     & Squared exponential kernel  (\ref{eq:SE})  & Cauchy kernel  (\ref{eq:CF})  \\
\hline
CC    &   $ 0.651\pm 0.194 \pm  8.29\%  $  & $0.641 \pm 0.191\pm 8.30\%  $   \\
CC+ CL   & $0.654 \pm 0.156 \pm 6.90\% $    &  $ 0.660\pm 0.161 \pm 7.45\% $  \\

Pantheon+ $H_{0}^{\text{R}}$ + CC+ CL    &$ 0.708\pm0.052 \pm 3.40\% $     &$ 0.705\pm 0.054 \pm 3.34 \%$    \\
Pantheon+ $H_{0}^{\text{HW}}$ + CC+ CL    &    $ 0.712\pm 0.054 \pm 3.46\%$  &$0.709 \pm 0.055 \pm 3.41\%$      \\
Pantheon+ $H_{0}^{\text{Planck}}$ + CC+ CL   & $0.663 \pm 0.053  \pm 2.15\% $    & $ 0.644\pm0.054 \pm 2.33\%$    \\
Pantheon + $H_{0}^{\text{TRGB}}$ + CC+ CL  &  $ 0.664 \pm 0.056 \pm 2.67\% $     &$ 0.668 \pm 0.059  \pm 2.73 \%$         \\
\hline
\end{tabular}
\caption{Obtained values for the density parameter $\Omega_{\text{eff}}^{0}$ with their corresponding 1$\sigma$ uncertainties.  The first column shows the data sets used to perform each reconstruction. The second column shows the obtained value for  $\Omega_{\text{eff}}^{0}$ using the 
squared exponential  kernel. The third column  shows the obtained value  for  $\Omega_{\text{eff}}^{0}$ using the Cauchy kernel.  The first value in  the second and third  columns represents the mean value of  $\Omega_{\text{eff}}^{0}$, the second value is the  1$\sigma$ statistical uncertainty and the third value is the numerical error associated with the calculation of $\Omega_{\text{eff}}^{0}$.}
\label{table:31}
\end{table}
\\\\
Using the constraint equation $\Omega_{m}^{0}+\Omega_{\Lambda}^{0}=1$, and the reconstruction that minimizes the $\chi^{2}$ statistic in Table \ref{table:3}, it can be shown that 
\begin{eqnarray}
    &&\Omega_{m}^{0}=0.336 \pm 0.0563 \pm 2.67\%, \quad \rightarrow  \quad \text{Squared Exponential}, \\
    &&\Omega_{m}^{0}= 0.332 \pm 0.0594  \pm 2.73\%, \quad \rightarrow   \quad \text{Cauchy function},
\end{eqnarray}
both values coincide with the $\Omega_{m}^{0}$ reported by the Planck 2018 collaboration, so there is not a tension effect on this parameter. Although this is  due to the size of the 1$\sigma$ uncertainties in   $\Omega_{\text{eff}}^{0}$.

%%%%%%%%%%%%%%%%%%%%%%%%%%%%%%%%%%%%%%%%%%%%
%%%%%%%%%%%%%%%%%%%%%%%%%%%%%%%%%%%%%%%%%%%%

\section{Updated constraints on dark energy models from Horndeski Cosmology}
\label{sec:Horn}

In the last few years, Horndeski theory of gravity~\cite{kobayashi2019horndeski} 
has been studied for scalar-field dark energy  \cite{zumalacarregui2020gravity,bellini2016constraints,garcia2020theoretical,Bayarsaikhan_2020,Frusciante:2018aew} where it has been show that its solutions can be ghost-free \cite{Helpin_2020} and undergo tachyonic instability roles \cite{Frusciante:2018vht}. Also, Galileon cosmology from Horndeski theory has been done in Refs.~\cite{Chow_2009,Peirone_2018,DeFelice:2010nf,Kennedy:2018gtx}. Furthermore, extensions up to quartic order have been considered in  Ref.~\cite{Deffayet_2009,Barreira_2013}.

The Horndeski theory is the most general  covariant scalar-tensor theory with a single scalar field whose Lagrange equation has derivatives of the metric and the scalar field up to second order.  The action of this theory is given by

\begin{equation}
S[g_{ab},\phi]=\int d^{4}x\sqrt{-g} \left [ \sum^{5}_{i=2}  \frac{1}{8 \pi G }   \mathcal{L} [g_{ab},\phi] +\mathcal{L}_{m} [g_{ab}] \right ], \label{eq:actionH}
\end{equation}
where the four Lagrangians $ \mathcal{L}_{i} $, are given by:
\begin{eqnarray}
    \mathcal{L}_{2}&=&G_{2}(\phi,X),\\
    \mathcal{L}_{3}&=&-G_{3}(\phi,X) \square \phi,  \\
    \mathcal{L}_{4}&=&   G_{4}(\phi,X)R+ G_{4X}\left [ (\square\phi)^{2}-\phi_{;ab}\phi^{;ab} \right ], \\
    \mathcal{L}_{5}&=&   G_{5}(\phi, X) G_{ab}\phi^{;ab}-\frac{G_{5X}(\phi,X)}{6}\left [ (\square \phi)^{3}+2\phi_{;\mu}^{\nu}\phi_{;\nu}^{\alpha}\phi_{;\alpha}^{\mu}-3  \phi_{;ab} \phi^{;ab} \square \phi \right ],
\end{eqnarray}
with $ G_{2} $, $ G_{3} $, $ G_{4} $ and $ G_{5} $ arbitrary functions of  the scalar field $ \phi $ and its kinetic term $2X=-\partial_{\mu} \phi\partial^{\mu}\phi$. In the above equations $ R $ the Ricci scalar, $ G_{\mu \nu} $ the Einstein tensor. We use the subscripts $ X $ and $\phi $ to denote partial derivatives, with respect to $ X $ and $ \phi $. The $ G_{i} $ functions are completely arbitrary, however, in the next section  will be explained how we have constrained these functions. 

The  detection of the gravitational waves GW170817 \cite{Abbott_2017}, the Gamma Ray Burst GRB 170817A and the a transient Swope Supernova Survey SSS17a \cite{Abbott_2017,Coulter_2017} places a strong constraint on the speed at which gravitational waves propagate \cite{Baker_2017,Creminelli_2017,Ezquiaga_2017}
\begin{equation}
-3\times 10^{-15}<c_{GW}-1<7\times 10^{-16}.    
\end{equation}
This  constraint motivates  to only consider only a subclass of Horndeski models, that is, those that meet the condition $c_{GW} = 1 $,
these models \cite{kobayashi2019horndeski} are given by  $ G_{4X} = 0$ and $G_{5} = 0$. The  Lagrangian for this models has the form 
\begin{equation}
 \mathcal{L}= G_{2}(\phi,X)-G_{3}(\phi,X) \square \phi+ G_{4}(\phi)R.\label{eq:Lagrangian}
\end{equation}
We analyze  the Lagrangian (\ref{eq:Lagrangian}) using a spatially flat,  homogeneous and isotropic  Friedmann-Lemaitre-Robertson-Walker  metric, $ds^{2}=dt^{2}-a(t)^{2}(dx^{2}+dy^{2}+dz^{2}  )$, in addition to describe the matter content of the Universe, we use  the energy-momentum tensor $T_{\mu\nu}$ for a perfect fluid
$T_{\mu \nu}=(\rho+p)u_{\mu}u_{\nu}+pg_{\mu \nu} \label{eq:3_3},$
and since in this work we are focusing  on cosmic late-time aspects, the density of radiation can be neglected, then the Friedmann equations are given by
\begin{eqnarray}
   &&  H^{2} = \frac{8 \pi G}{3}\rho_{m}+\varepsilon \label{eq:aa1},   \\  
    && \dot {\varepsilon}+3H(\varepsilon +P)=0,  
    \end{eqnarray}\label{eq:pr}
    where
\begin{eqnarray}
\varepsilon&=&\frac{1}{3}  \left [-G_{2}+2X(G_{2X}-G_{3\phi})  \right ]+2H\dot{\phi}\left [  -G_{4\phi}+ X G_{3X}\right ]+H^{2}\left [         1-2G_{4} \right ], \\
P&=&\frac{2}{3}\dot{H}[1-2G_{4}   ]-H^{2}(1-2G_{4})      +\frac{2}{3}(H\dot{\phi} +\ddot{\phi})\left  [  G_{4\phi}  -XG_{3X} \right ]\nonumber\\ &&
+\frac{1}{3}\left [ G_{2}-2X( G_{3\phi} -2G_{4\phi \phi}  ) \right ]+\frac{2H\dot{\phi}}{3}\left [ G_{4\phi}+X  G_{3X} \right ]\label{eq:pr2}.
\end{eqnarray}
From this point forward we will analyse two different cases for the $G_{i}$'s using the potential 
\begin{equation}
 V=C\exp(n\phi),  \label{eq:potential} 
\end{equation}
with  $n$ and $C$  free parameters.

%%%%%%%%%%%%%%%%%%%%%%%%%%%%%%%%%%%%%%%%%%%%
%%%%%%%%%%%%%%%%%%%%%%%%%%%%%%%%%%%%%%%%%%%%

\subsection{Quintessence}
In this first example we will analyse the model 
\begin{align}
 G_ {2}(\phi, X) =  X-V (\phi), \quad  G_ {3} = 0, \quad    G_ {4} = 1/2,    \label{eq:g1}
\end{align}
which is a Horndeski model equivalent to  a Quintessence model \cite{bellini2020hi_class}. We can obtain the field equations for this model  if we substitute  (\ref{eq:g1}) in (\ref{eq:aa1})-(\ref{eq:pr2}) to obtain 
\begin{eqnarray}
     &&H^{2} = \frac{8 \pi G}{3}\rho_{m}+\varepsilon, \label{eq:manipulando}  \\  
     &&\dot {\varepsilon}+3H(\varepsilon +P)=0\label{eq:KleinG2}, 
    \end{eqnarray}
    where
\begin{eqnarray}
    && \varepsilon = \frac{1}{3} [- G_{2} +        2XG_{2X}     ]\label{eq:DensQ},\\
    && P=\frac{1}{3}[   G_{2}   ]\label{eq:PressQ}.
    \end{eqnarray}
In this model, the effective equation of state for the scalar field is given by
\begin{equation}
\omega_{\phi}=\frac{\frac{1}{2}\dot{\phi}^{2}- V(\phi)}{\frac{1}{2}\dot{\phi}^{2}+ V(\phi)},
\end{equation}
which in the limit where $\dot{\phi}\approx 0$, $\omega_{\phi}\approx-1.$
Since we need that our model reduces  to $ \Lambda$CDM at least for $ z = 0 $, it is necessary that $ \dot {\phi} (z = 0) = 0 $, which represents  an initial condition for the scalar field $\phi$.  We can obtain the value of $C$ from (\ref{eq:potential}) by evaluating (\ref{eq:manipulando}) at $ z = 0 $
\begin{equation}
    V(\phi(z=0))=3H_{0}^{2}[1-\Omega_{m}^{0}]\equiv C\label{eq:Cond}.
\end{equation}
Substituting  (\ref{eq:DensQ}) and (\ref{eq:PressQ}) in the continuity equation  (\ref{eq:KleinG2}), it is possible to show that the scalar field must satisfy  the Klein-Gordon (KG) equation
\begin{equation}
    \ddot{\phi}+3H\dot{\phi}+V_{\phi}=0. \label{eq:kleinc}
\end{equation}
It is useful to change the KG from the cosmic time  to redshift:
 \begin{equation}
     \phi''[H(1+z)]^{2}+\phi'[H'(1+z)+H](1+z)H-3H^{2}(1+z)\phi'+V_{\phi}=0\label{ec:rec}.
\end{equation}   
This equation requires two initial conditions: the first one was obtained  previously, i.e $ \dot {\phi} (z = 0) = 0 $ and the second changes depending on the potential that we choose, specifically for potentials of the type (\ref{eq:potential}) must satisfy $\phi(z=0)=0$. To obtain the value of $ n $ that best fits the reconstructions, we minimize the following quantity
\begin{equation}
\chi^{2}= \sum_{i}^{N_{\text{recons}} }\frac{    \left (  H_{\text{recons}}(z_{i})- H_{\text{Q}}(z_{i};n) \right )^{2}    }{\sigma_{\text{recons}}(z_{i})^{2}}, \label{eq:cQ}
\end{equation}
where $N$ represents the number of points in the reconstruction, $H_{\text{recons}}(z_{i})$ is the Hubble parameter of the reconstruction evaluated in the redshift $z_{i}$,   $\sigma_{\text{recons}}(z_{i})$ is the  1$\sigma$ uncertainty  of the  reconstruction and  $H_{\text{Q}}(z_{i};n)$  is the Hubble parameter for Quintessence. Notice that the value of $C$ can be obtained by considering the condition (\ref{eq:Cond}) evaluate at $z=0$, therefore our free parameter will be $n$. For the squared exponential kernel we obtain $n=1.0\times 10^{-9}$, while for the Cauchy  kernel we find $n=1.1\times 10^{-9}$.

\begin{figure}
    \centering
    \includegraphics[scale=0.34]{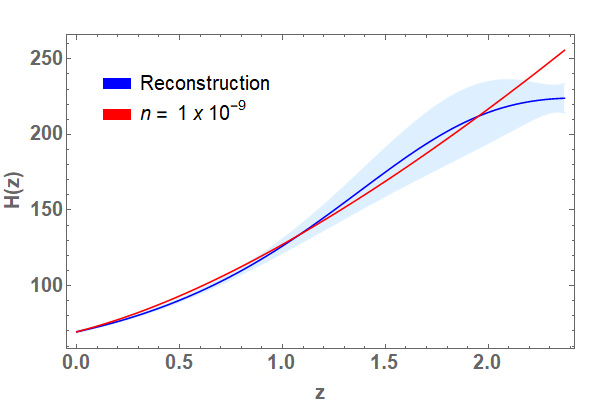} 
\includegraphics[scale=0.34]{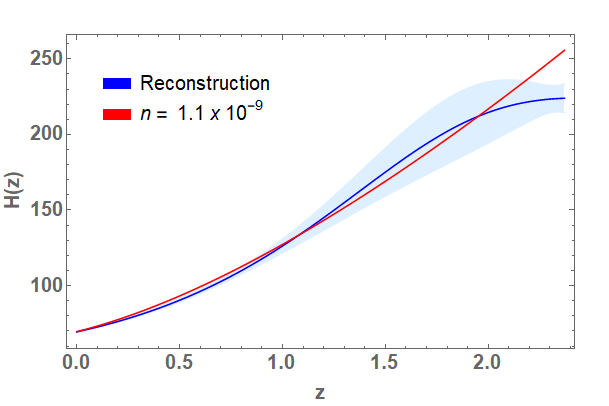} 
    \caption{Evolution of $ H (z) $ obtained  by solving (\ref{ec:rec}) with the potential $ V = C \exp (n \phi) $, compared with the $ H (z) $ obtained from the reconstruction with the squared exponential (left) and  Cauchy (right) kernels  at 2$\sigma $ level. The values of $n$ are obtained minimizing  (\ref{eq:cQ}).}
    \label{fig:Quintessence}
\end{figure}

In Fig. \ref{fig:Quintessence} we show the evolution of $ H (z) $ by solving (\ref{ec:rec}) for  the potential (\ref{eq:potential}) with the mean value of $H_{0}$ obtained from the  table \ref{table:1} and the  mean value of $\Omega^{0}_{m}$ obtained from the table \ref{table:2}, that  correspond to the reconstruction Pantheon + $H^{ \text{TRGB}}_{0}$ + CC + CL. This result is compared with the Hubble parameter obtained from the reconstruction named Pantheon + $H^{ \text{TRGB}}_{0}$ + CC + CL.

%%%%%%%%%%%%%%%%%%%%%%%%%%%%%%%%%%%%%%%%%%%%
%%%%%%%%%FIGURES QUINTESSENCE %%%%%%%%%%%%%%
%%%%%%%%%%%%%%%%%%%%%%%%%%%%%%%%%%%%%%%%%%%%
Notice that the Quintessence  model is within  the 2$\sigma$ confidence intervals for the reconstructions on $0<z<2$, however for $z>2$, this model cannot describe the reconstructions on this confidence interval. 

%%%%%%%%%%%%%%%%%%%%%%%%%%%%%%%%%%%%%%%%%%%%
%%%%%%%%%%%%%%%%%%%%%%%%%%%%%%%%%%%%%%%%%%%%

\subsection{K-essence}
\label{sub:G33}

In this section we take the model
 
\begin{equation}
  G_{2}=X-V,\quad  G_{3}=-c_{1}V,\quad  G_{4}=1/2,\label{eq:g2}
\end{equation}
where $ c_ {1} $ is a positive constant  with units of $ s^{2} $,  in particular, we choose $ c_ {1} = 1 s^{2}$.  If  $G_3$ is an increasing function on the redshift, then  values of $c_1$ close to 1$s^{2}$  tend to decrease the expansion rate of the universe with respect the expansion rate given by $\Lambda$CDM. Meanwhile, for large values of $c_{1}$ our model goes to  $\Lambda$CDM.  Notice that this model can be reduced to a K-essence model by performing an integration by parts of the corresponding action (\ref{eq:actionH}), where the term $G_3$ vanishes\footnote{See Equation (2.2) from \cite{Deffayet:2010qz}.}:
%\textbf{Integrating by parts the  corresponding action of the model it can be shown that the ac\cite{Deffayet:2010qz}}
\begin{equation}
  G_{2}=X-V-2Xc_{1}V_{\phi},
  %\quad G_{3}= 0,
  \quad  G_{4}=1/2,\label{eq:g2Correction}.
\end{equation}
 The  Friedmann equations for this model are  given by
\begin{eqnarray}
     && H^{2} = \frac{8 \pi G}{3}\rho_{m}+\varepsilon, \label{eq:ecx}  \\  
     && \dot {\varepsilon}+3H(\varepsilon +P)=0,    \label{eq:res}
    \end{eqnarray}
    with
\begin{eqnarray}
&& \varepsilon=\frac{1}{3}[V+X-2XV_{\phi}c_{1}], \\
&& P=\frac{1}{3}[X-V-2XV_{\phi}c_{1}].
\end{eqnarray}
Therefore, the equation of state for the scalar field has the form
\begin{align}
    \omega_{\phi}=\frac{X-V-2XV_{\phi}c_{1}}{V+X-2XV_{\phi}c_{1}}. \label{eq:ax}
\end{align}
If $\dot{\phi}(z=0)=0$, (\ref{eq:ax}) reduces to 
   $ \omega_{\phi}=-1$.
We can obtain the initial condition for $ \phi $ if we evaluate (\ref{eq:ecx}) at $ z = 0 $. In particular, for the potential $ V = C \exp (n \phi) $, it is necessary that $ \phi (z = 0) = 0 $, with $ C = 3H_ {0} ^ {2} [1 -\Omega_{m}^{0}] $. To obtain the evolution of $ \phi (z) $, it is useful to rewrite (\ref{eq:res}) as
\begin{equation}
    -\frac{1}{3}\frac{\mathrm{d} }{\mathrm{d} z}\left [ V+H^{2}(1+z)^{2}\phi'^{2}(\frac{1}{2}-V_{\phi}c_{1})\right ]+(1+z)\left [1-2 V_{\phi}c_{1}  \right ]H^{2}\phi'^{2}=0. \label{eq:rec2}
\end{equation}
To obtain the value of $ n $ that best fits the reconstructions, we minimize the following quantity
\begin{equation}
\chi^{2}= \sum_{i}^{N_{\text{recons}} }\frac{    \left [ (  H_{\text{recons}}(z_{i})- H_{\text{EQ}}(z_{i};n) \right ]^{2}    }{\sigma_{\text{recons}}(z_{i})^{2}}, \label{eq:ceq2}
\end{equation}
where $ N $ represents the number of points in the reconstruction, $ H_ {\text{recons}} (z_ {i}) $ is the Hubble parameter of the reconstruction evaluated at redshift $ z_ {i} $, $ \sigma_ {\text {recons}} (z_ {i}) $ is the 1$\sigma $ uncertainty of the reconstruction and $ H_{\text {EQ}}(z_{i};n) $ is the Hubble parameter for  K-essence. For the squared exponential kernel we get $n=79.31$ while for the  Cauchy kernel  $n=77.41$.

\begin{figure}
    \centering
    \includegraphics[scale=0.35]{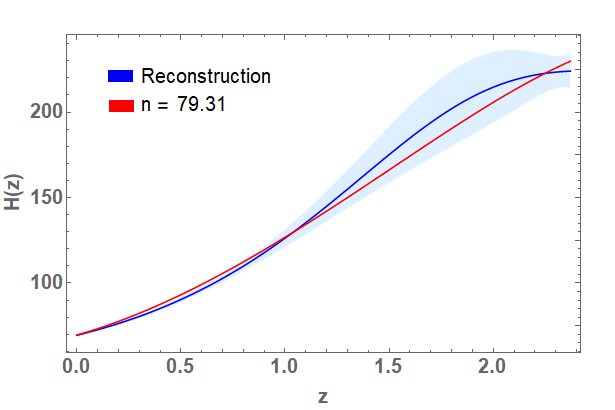} 
\includegraphics[scale=0.35]{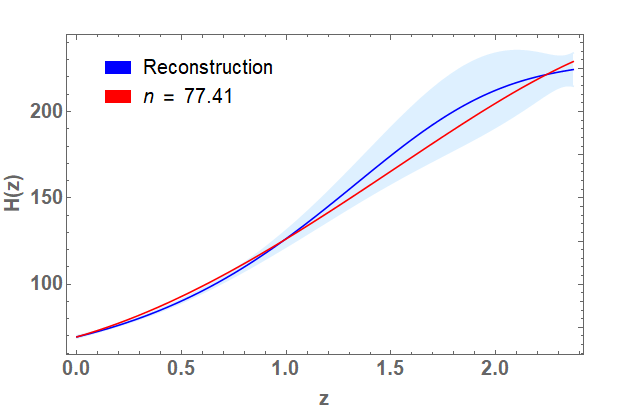}
    \caption{ Evolution of $ H (z) $ obtained from (\ref{eq:rec2}) with the potential $ V = C \exp (n \phi) $, compared with the $ H (z) $ obtained from the reconstruction with the squared exponential (left) and  Cauchy (right)      kernels  at 2$\sigma $ level. The values of $n$ are obtained minimizing the equation (\ref{eq:ceq2}).}
    \label{fig:G3}
\end{figure}

In Fig. \ref{fig:G3} we show the evolution of the Hubble parameter obtained  by solving  (\ref{eq:rec2}) for  the potential (\ref{eq:potential}) with the mean value  of $H_{0}$ and the mean value of $\Omega^{0}_{m}$ obtained from the reconstruction carried out with the Pantheon + $H^{ \text{TRGB}}_{0}$ + CC + CL data. This result is compared with the $H (z)$ obtained from the reconstruction carried out with the data of Pantheon + $H^{\text{TRGB}}_{0}$ + CC + CL. 

\begin{table}
\centering
\begin{tabular}{  |p{4.5cm} |p{5.5cm} |p{4.2cm} |}
\hline
\multicolumn{3}{ |c| }{Reduced $\chi^2$ at 1$\sigma$} \\
\hline
Model     & Squared exponential kernel  (\ref{eq:SE})  & Cauchy kernel  (\ref{eq:CF})  \\
\hline
Quintessence    &   $ 9.73  $  & $8.47$   \\
K-essence   & $7.78 $    &  $6.70$  \\
\hline
\end{tabular}
\caption{ Reduced $\chi^{2}$-statistics between the Pantheon + $H^{ \text{TRGB}}_{0}$ + CC + CL reconstruction  and the Horndeski model
using the squared exponential and Cauchy kernel, respectively.}
\label{table:CRE1}
\end{table}

\begin{table}
\centering
\begin{tabular}{ |p{4.5cm} |p{5.5cm} |p{4.2cm} |}
\hline
\multicolumn{3}{ |c| }{Reduced $\chi^2$ at 2$\sigma$} \\
\hline
Model     & Squared exponential kernel  (\ref{eq:SE})  & Cauchy kernel  (\ref{eq:CF})  \\
\hline
Quintessence    &   $ 2.43  $  & $2.11$   \\
K-essence   & $1.94 $    &  $1.67$  \\
\hline
\end{tabular}
\caption{Reduced $\chi^{2}$-statistics between the Horndeski model and the Pantheon + $H^{ \text{TRGB}}_{0}$ + CC + CL reconstruction  taking the uncertainties at the two sigma interval using the squared exponential and Cauchy kernel, respectively.}
\label{table:CRE2}
\end{table}
Table \ref{table:CRE1} shows a comparison between  the value of the reduced $\chi^{2}$-statistic for each model. It is observed that both models have a high  value of  the reduced $\chi^{2}$-statistics, this may be due to a possible overfitting of the reconstructions to the observations  in the region $0 < z <1$, this results are reported in Table \ref{table:CRE2} as a comparison of the reduced $\chi^{2}$-statistic assuming the value of the uncertainties of the reconstruction at 2$\sigma$ level. It is observed that in this case we have a reasonable value for the reduced $\chi^{2}$-statistic for the K-essence model.

%%%%%%%%%%%%%%%%%%%%%%%%%%%%%%%%%%%%%%%%%%%%
%%%%%%%%%FIGURES HORNDESKI G3 %%%%%%%%%%%%%%
%%%%%%%%%%%%%%%%%%%%%%%%%%%%%%%%%%%%%%%%%%%%

%%%%%%%%%%%%%%%%%%%%%%%%%%%%%%%%%%%%%%%%%%%%
%%%%%%%%%%%%%%%%%%%%%%%%%%%%%%%%%%%%%%%%%%%%

\section{Conclusions}
\label{sec:conclu}

In this paper we improved reconstructions of the late time expansion of the universe using Gaussian Processes. We show the possibility of using our  reconstructions to study the Hubble tension and to obtain constrains on the Galileon $G_2$ from Horndeski theory of gravity. Our work presents an improvement  over previous methods reported in the literature, and we describe that the use of compressed data sets to perform  Gaussian Processes is not the best option, since the methods that use  this type of  data sets such as  the  \textit{Covariance functions reconstruction method} and  the \textit{Reconstruction with weighted polynomial regression method}  have uncertainties  in $ H_ {0} $ twice larger than ours.

We showed that the obtained values for $ H_ {0} $ from each reconstruction reach differences up to 13$\sigma $ depending on the prior used to calibrate the SNeIa data given by Pantheon. Using the $ \chi^{2}$-statistics, we found that the reconstruction that best fit the observations is the combination Pantheon + $ H ^ {\text{TRGB}}_ {0}$ + CC + CL, with Pantheon dataset calibrated using $ H_ {0} $ measured by The Carnegie-Chicago Hubble Program \cite{freedman2019carnegie}. Our results reports that the $H_ {0}$ value that best describes the observations of the late universe is $ H_{0} = 69.53 \pm 0.37 $, and this value links a tension of 3.4$\sigma$ with respect the value of $\Lambda$CDM inferred from Planck 2018 data. However, it is still possible to improve the quality of the reconstructions presented in this work by recalibrating the Clustering of Galaxies sample.

Furthermore, assuming a parametric form for the dark energy equation of state we obtain new constraints on the density parameter for this exotic fluid at the present time. 
By defining  (\ref{eq:estimador}) we derive a way to estimate the numerical error associated with the $\Omega_ {\text{eff}}^{0}$ calculation. We  also show that the density parameters obtained from the reconstructions are not in tension with those obtained by the Planck 2018 collaboration assuming a flat $\Lambda$CDM, although this is mainly due to the size of our 1$\sigma$ uncertainties.

Finally, we showed that Horndeski models as Quintessence type under the potential (\ref{eq:potential}) are not able to fit  the reconstructions of the Hubble flow within the  2$\sigma$ confidence interval, therefore there is a possibility that these models are not good candidates to describe the cosmic late time observations. The K-essence model discussed in Sec.\ref{sub:G33} with an exponential-type potential fits reconstructions at 2$\sigma$ confidence interval and therefore represents a good candidate to describe the late time expansion. Furthermore, in contrast with the results presented in \cite{Garc_a_Garc_a_2020},  where an exponential potential is analysed\footnote{Also so-called \textit{Modulus model.}} with a fixed positive $n$ prior value, in our analysis we obtained  best fit positive values for $n$ in both Quintessence and K-essence cases.

It is expected  that the K-essence model  modify the history of the large-scale structure formation, and therefore it will be interesting to analyse  the first order linear perturbations of this model  since these will tell us if it is also a candidate to solve the $ S_ {8} $ tension \cite{di2020cosmology}. These analyses are an ongoing project, and they will be reported elsewhere.

%%%%%%%%%%%%%%%%%%%%%%%%%%%%%%%%%%%%%%%%%%%%
%%%%%%%%%%%%%%%%%%%%%%%%%%%%%%%%%%%%%%%%%%%%

\acknowledgments

CE-R acknowledges the Royal Astronomical Society as FRAS 10147. CE-R and M.R are supported by DGAPA-PAPIIT-UNAM Project IA100220. This article is also based upon work from COST action CA18108, supported by COST (European Cooperation in Science and Technology). The simulations were performed in Centro de c\'omputo Tochtli-ICN-UNAM. The authors thank the anonymous reviewer whose comments/suggestions helped improve and clarify this manuscript.

%%%%%%%%%%%%%%%%%%%%%%%%%%%%%%%%%%%%%%%%%%%%
%%%%%%%%%%%%%%%%%%%%%%%%%%%%%%%%%%%%%%%%%%%%
\bibliographystyle{unsrt}
\bibliography{refs}

\end{document}